\documentclass[12pt]{article}
\pdfoutput=1
\usepackage{graphicx}
\usepackage{epstopdf}
\usepackage{amsmath}
\usepackage{amsfonts}
\usepackage{amssymb}
\usepackage{color}
\usepackage{mathrsfs}
\usepackage{breqn}

\newcommand{\be}{\begin{equation}}
\newcommand{\ee}{\end{equation}}
\newcommand{\barr}{\begin{array}}
\newcommand{\earr}{\end{array}}

\usepackage{latexsym}
\usepackage{amsthm}
\usepackage{amsmath}
\usepackage{graphicx}
\usepackage{amssymb}
\usepackage{bbm}
\newcommand{\gsim}{\lower.7ex\hbox{$\;\stackrel{\textstyle>}{\sim}\;$}}
\newcommand{\lsim}{\lower.7ex\hbox{$\;\stackrel{\textstyle<}{\sim}\;$}}

\newcommand{\bea}{\begin{eqnarray}}
\newcommand{\eea}{\end{eqnarray}}

\newcommand{\comment}[1]{}

\def\e{\mathrm{e}}

\def\Mp{M_{\rm pl}}

\newcommand{\nn}{\nonumber}

\def\p{\partial}

\def\({\left(}
\def\){\right)}
\def\[{\left[}
\def\]{\right]}
\def\e{\begin{equation}}
\def\q{\end{equation}}
\def\m{\begin{eqnarray}}
\def\n{\end{eqnarray}}

\begin{document}

%\begin{titlepage}

\setcounter{page}{1} \baselineskip=15.5pt \thispagestyle{empty}

\begin{flushright}
%hep-th/yymmnnn\\
\end{flushright}
\vfil

\begin{center}

{\Large \bf The four-point correlation function of graviton during inflation}
\\[0.7cm]
{Tian-Fu Fu$^{\heartsuit,\ \Diamond}$ \footnote{futianfu@itp.ac.cn}  and Qing-Guo Huang$^\heartsuit$ \footnote{huangqg@itp.ac.cn}}
\\[0.7cm]
%\vspace{.7cm}
%\vspace{.3cm}

{\normalsize { \sl $^\heartsuit$ State Key Laboratory of Theoretical Physics, Institute of Theoretical Physics, \\ Chinese Academy of Science, Beijing 100190, China}}\\
\vspace{.2cm}

{\normalsize { \sl $^\Diamond$  University of the Chinese Academy of Sciences, Beijing 100190, China}}
\vspace{.3cm}

\end{center}

\vspace{.8cm}

\hrule \vspace{0.3cm}
{\small  \noindent \textbf{Abstract} \\[0.3cm]
We expand the Einstein-Hilbert action with a positive cosmological constant up to the fourth order in terms of gravity fluctuations, and then use the in-in formalism to calculate the four-point correlation function for gravitational waves, including both contact and exchange diagrams, generated during a period of exactly de Sitter expansion. In addition, we also present the general properties of the $n$-point function of graviton in terms of both circularly and linearly polarized states.
}
 \vspace{0.3cm}
\hrule

%\vfil
\begin{flushleft}
%\today
%March 20, 2008
\end{flushleft}

\vspace{8cm}

\newpage
%%%%%%%%%%%%%%%%%%%%%%%%%%%%%%%%%%%%%%%
\section{Introduction}

The discovery of the accelerating expansion of the Universe through observations of distant supernovae \cite{Riess:1998cb,Perlmutter:1998np}, together with the theory of inflation \cite{Starobinsky:1980te,Guth:1980zm,Linde:1981mu,Albrecht:1982wi}, suggest the possibility that our universe approaches de Sitter geometries in both the far past and the far future.
However, what is the nature of dark energy and inflation is still one of the deepest puzzles in modern physics. We believe that the quantum theory of gravity need to be understood before solving this puzzle.

In an asymptotically flat spacetime a unitary and analytic S-matrix can be well-defined. Even though there is no notion of an S-matrix in asymptotically anti de Sitter spacetime, one has the correlation function for the graviton which is equivalent to the correlation function of the stress tensor in the conformal field theory on the boundary \cite{Maldacena:1997re}. Such a correspondence between gravity and conformal field theory is crucial for us to understand the quantum theory of gravity in the anti de Sitter space. %Similarly the correlation functions of graviton in de Sitter background may also shed light on the quantum gravity theory in de Sitter space.
On the other hand, tensor perturbations are also generated during inflation \cite{Starobinsky:1979}, whose direct detection would be taken as the evidence for inflation, and the non-Gaussianities of gravitational waves is a key feature of tensor perturbations as well as the scalar perturbations \cite{Maldacena:2002vr}.
The three-point correlation function of graviton in de Sitter background has been calculated in \cite{Maldacena:2011nz}. In this paper we will investigate the general properties of $n$-point correlation function of graviton, and explicitly compute the four-point function of graviton in de Sitter background. Even though these gravity correlation functions appear to be outside the reach of the experiments occurring in the near future, we hope that the correlation functions of graviton in de Sitter background may also shed light on the quantum gravity theory in de Sitter space.

\begin{figure*}[htbp]
\centering
\begin{center}
\includegraphics[scale=0.5]{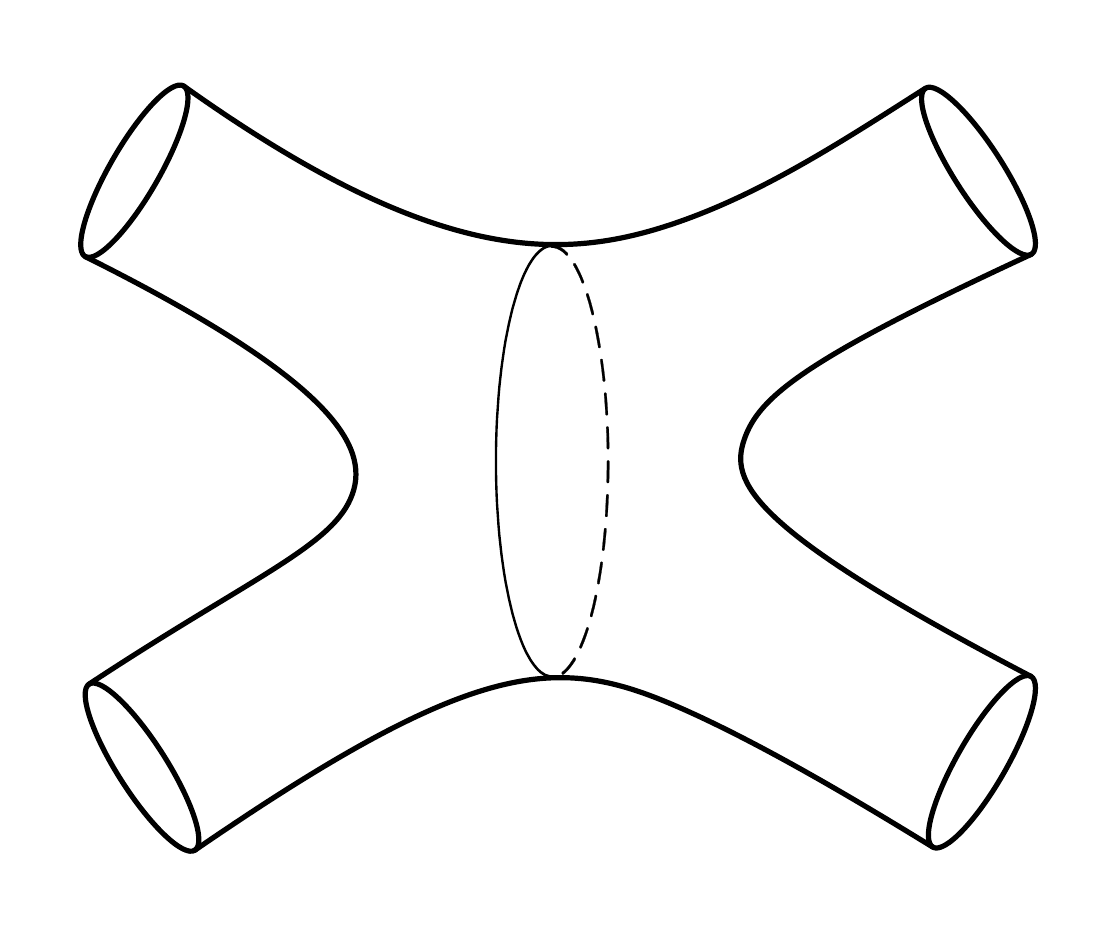}
\end{center}
\caption{The unique diagram of the four-point function of graviton at tree level in string theory. }\label{fig:string}
\end{figure*}

String theory is the most promising quantum theory of gravity. In principle, the diagram of the four-point function of graviton at tree level in string theory is unique (see Fig.~\ref{fig:string}).
Unfortunately, string theory cannot be well formulated in de Sitter space. So we do not know how to calculate the four-point function of graviton in de Sitter background in string theory. Up to now, general relativity (GR) proposed by Albert Einstein one hundred years ago is still the simplest theory of gravity which is consistent with all of the experimental data, and every consistent string theory must contain a massless spin-2 state whose interactions reduce at low energy to general relativity. Here we will expand the Einstein-Hilbert action with a positive cosmological constant up to the fourth order in terms of gravity  fluctuations, and utilize the in-in formalism to calculate the four-point correlation function of graviton, including both the contact and exchange diagrams, in de Sitter background.

This paper is organized as follows. In section \ref{section:action} we expand the Einstein-Hilbert action with a positive cosmological constant up to the fourth order in terms of gravitational waves. The parity of tensor perturbations and the general properties of $n$-point correlation function of graviton will be discussed in section \ref{section:parity}. The four-point function of graviton in de Sitter background will be computed explicitly in section \ref{section:4pt}. Summary and discussion will be given in section \ref{section:dis}. In addition, a brief review of tensor power spectrum and three-point function of graviton in de Sitter background is presented in Appendix \ref{app:3pt}, and we fix the representation of polarization tensors and clarify the general four-momentum configuration into $6$ parameters in Appendix \ref{app:poltensor}. The constraints on the correlation function of graviton from the de Sitter isometries are discussed in Appendix \ref{app:isometries}.

\section{The action up to fourth order in terms of tensor fluctuations in de Sitter background}
\label{section:action}

In this paper we focus on the Universe whose dynamics is govern by the Einstein-Hilbert action with a positive cosmological constant $\Lambda$, namely
\e
S=\frac{\Mp^2}{2}\int dtd^3x \sqrt{-g}(R-2\Lambda),
\label{EHaction}
\q
where $\Mp\equiv (8\pi G)^{-1/2}$ is the reduced Planck mass. The background spacetime is described by nothing but de Sitter with metric
\m
ds^2=-dt^2+e^{2Ht} dx_i dx_i,
\n
where $H$ is the Hubble constant related to $\Lambda$ by $H^2=\Lambda/3$.

During a period of exactly de Sitter expansion, there are only two physical degrees of freedom corresponding to the tensor fluctuations (gravitational waves). In order to make the dynamical degrees of freedom manifest, we call for the the ADM formalism \cite{Arnowitt:1962hi}, namely
\e
ds^2 = -N^2dt^2+g_{ij}(dx^i+N^idt)(dx^j+N^jdt),
\q
where $N$ and $N_i$ are lapses and shifts. Now the action (\ref{EHaction}) reads
\e
\label{action}
S=\frac{\Mp^2}{2}\int dtd^3x\sqrt{g_3}(NR^{(3)}-6NH^2+N^{-1}(E_{ij}E^{ij}-E^2)),
\q
where $R^{(3)}[g_{ij}]$ is the three-dimensional Ricci scalar associated with the metric $g_{ij}$, and the symmetric tensor $E_{ij}$ is defined by
\e
E_{ij}=\frac{1}{2}(\dot{g}_{ij}-\nabla_i N_j-\nabla_j N_i),
\q
here the dot denotes the derivative with respect to the cosmic time $t$, and $E\equiv g^{ij} E_{ij}$. Note that the extrinsic curvature is $K_{ij}= N^{-1}E_{ij}$. In this sense $E_{ij}E^{ij}-E^2$ can be taken as the``kinetic term" which governs the dynamics of the extrinsic curvature.
In this form the spatial coordinate reparameterization is explicitly symmetric, while the
time reparameterization is not so obvious since $N^i$ mixes up the space and time coordinates.
In the ADM form the metric $g_{ij}$ is dynamical, but $N$ and $N_i$ are Lagrange multipliers. It is convenient to choose a gauge for $g_{ij}$ in which the time and spatial reparametrizations are fixed, and then the metric perturbation can be defined by
\e
g_{ij}=a^2 \exp{(h)}_{ij},
\q
where $a=e^{Ht}$ and the gravity fluctuations are transverse and  traceless, i.e. $\p_i h_{ij}=0$ and $h_{ii}=0$. Such a symmetric transverse and traceless metric $g_{ij}$ encodes two physical degrees of freedom of gravitational waves in GR.

From Eq.~(\ref{action}), the equations of motion for $N$ and $N_i$ are given by
\m
R^{(3)}-6H^2 -N^{-2}\(E_{ij}E^{ij}-E^2\)&=&0,  \label{c1} \\
\nabla_i\left[N^{-1}\left(E^i_j-E\delta^i_j\right)\right]&=&0. \label{c2}
\n
In general, in order to work out the action up to fourth order in terms of gravitational waves, one only need to solve the above two equations up to second order in the ADM formalism because the third and fourth order perturbations multiplies the first and zeroth order constraint equation, respectively, which vanishes. See \cite{Maldacena:2002vr} for the discussion in detail. Therefore we decompose
\m
N&=&1+\alpha^{(1)}+\alpha^{(2)}+\cdots , \\
N_i&=&N_i^{(1)}+N_i^{(2)}+\cdots .
\n
Furthermore, the shift $N_i^{(n)}$ can be decomposed into its spin-0 scalar $\psi^{(n)}$ and spin-1 vector components $\beta^{(n)}_i$ in the form of
\e
N_i^{(n)}=\beta_i^{(n)}+\partial_i \psi^{(n)}.
\q
Here $\alpha^{(n)}, \beta_i^{(n)}, \psi^{(n)}\sim {\cal{O}}(h^n) $. Now one can solve Eqs.~(\ref{c1}) and (\ref{c2}) order by order. Up to order of ${\cal{O}}(h^2) $, we get
\m
&&\alpha^{(1)}=\alpha^{(2)}=\beta_i^{(1)}=\beta_i^{(2)}=\psi^{(1)}=0, \label{constraint1} \\
&&\partial_i\partial_i \psi^{(2)}=-\frac{1}{16 H}(h^{\prime}_{ij} h^{\prime}_{ij}+ \partial_k h_{ij}\partial_k h_{ij} ), \label{constraint2}
\n
where the prime $\prime$ denotes the derivative with respect to the conformal time $\eta\equiv\int^t dt^\prime/a(t^\prime)$.

Plugging the solutions in \eqref{constraint1} and \eqref{constraint2} into action \eqref{action} and  expanding it up to the fourth order, we obtain
\e
S= S^{(2)}+S^{(3)}+S^{(4)},
\q
 where
 \begin{eqnarray}\label{2nd-action}
 S^{(2)}&=&\frac{\Mp^2}{8}\int d\eta d^3x a^2(h^\prime_{ij}h^\prime_{ij}-\partial_l h_{ij}\partial_l h_{ij}),
 \\ \label{3rd-action}
 S^{(3)}&=&\frac{\Mp^2}{4} \int d \eta d x^3  a^2(c_1 h_{ik}h_{jl}-\frac{c_2}{2} h_{ij} h_{kl})\partial_k \partial_l h_{ij},
 \\ \label{4th-action}
 S^{(4)}&=& \Mp^2 \int d\eta dx^3\left[-\frac{d_1}{4} a (\partial_i \psi^{(2)})h_{jk}^{\prime}\partial_i h_{jk}
+\frac{d_2}{48} a^2 h^{\prime}_{ij}\(h^{\prime}_{jk}h_{kl}-h_{jk}h^{\prime}_{kl}\)h_{li} \right.
\nonumber\\ && \left.
+\frac{d_3}{48} a^2 \partial_m h_{ij}\(h_{jk}\partial_m h_{kl}-\partial_m h_{jk}h_{kl}\)h_{li}
+d_4 a^2 \partial_k h_{ij} \partial_l h_{ij} h_{km}h_{lm} \right.
\nonumber\\  && \left.
+\frac{d_5}{8} a^2 h_{ij}\(-\frac{1}{2}\partial_i h_{jk}\partial_m h_{kl}+\partial_m h_{jk}\partial_i h_{kl}\) h_{lm}
\right],
 \end{eqnarray}
Here the coefficients $c_i(=1)$ and $d_i (=1)$ are used to denote the different interaction terms in the third and fourth order action.
 %The reason of using these seemingly redundant coefficients is to track the contribution to correlation function from different parts of action.

%\section{Review of computation of spectrum and bispectrum }
\section{Parity of tensor perturbations and the general properties of $n$-point correlation function of graviton}
\label{section:parity}

According to the quadratic action in Eq.~(\ref{2nd-action}), the linear perturbation equation can be solved in the Fourier space. So one can decompose the perturbations into momentum modes as follows
\e
h_{ij}(\eta,\mathbf{x})=\int \frac{d^3k}{(2\pi)^3}\tilde{h}_{ij}(\eta,\mathbf{k})e^{i\mathbf{k}\cdot\mathbf{x}},
\label{dech}
\q
where
\m
 \tilde{h}_{ij}(\eta,\mathbf{k})&=&\sum_{s}e^s_{ij}(\mathbf{k}) h^{s}_{\mathbf{k}}(\eta),
\n
$e_{ij}^{s}(\mathbf{k})$ is the polarization tensor which has the following properties
\m
e_{ij}^{s}(\mathbf{k})=e_{ji}^{s}(\mathbf{k}),\quad e_{ii}^{s}(\mathbf{k})=k_je_{ij}^{s}(\mathbf{k})=0.
\n
For convenience, the polarization tensor is normalized as
\m
e^{s_1}_{ij}(\mathbf{k})e^{s_2*}_{ij}(\mathbf{k})=4\delta_{s_1 s_2}.
\n
Usually one can adopt the linear or circular polarization.

\subsection{Circular polarization}

In the literatures the circular polarization is widely used because they are the helicity eigenstates of gravitational waves.
For the circular polarization, the polarization tensor takes the form
\e \label{CPtensor}
e^{+2}_{ij}(\mathbf{k})=
\left(
  \begin{array}{ccc}
    1 & i & 0\\
    i & -1 & 0\\
    0 & 0 & 0\\
  \end{array}
  \right), \quad
  e^{-2}_{ij}(\mathbf{k})=
\left(
  \begin{array}{ccc}
    1 & -i & 0\\
    -i & -1 & 0\\
    0 & 0 & 0\\
  \end{array}
  \right),
 \q
where $\pm 2$ denote two circularly polarized states which propagate along $z$ direction. From the above formula, the circular polarization tensors satisfy
\m
e^\lambda_{ij}(-\mathbf{k})=e^{-\lambda}_{ij}(\mathbf{k})=(e^\lambda_{ij}(\mathbf{k}))^*,
\n
where $\lambda=\pm 2$.
The most general solution of quantized tensor perturbation can be written by
 \m
 h^{\lambda}_{\mathbf{k}}(\eta)&\equiv& \psi_k a_\lambda(\mathbf{k})+\psi^*_k a_\lambda^\dagger(-\mathbf{k}),
 \label{cphsk}
 \n
where $\psi_k$ is the mode function which satisfies
\e
 \psi_k^{\prime\prime}-\frac{2}{\eta}\psi_k^\prime+k^2\psi_k=0,
\q
and $k=|\mathbf{k}|$.
After choosing the standard Bunch-Davies vacuum \cite{Bunch:1978yq}, the solution of mode function is given by
 \e\label{modefunction}
 \psi_{k}(\eta)=\frac{H}{\sqrt{2k^3}}e^{-ik\eta}(1+ik\eta)~.
 \q
Now the creation and annihilation operators are normalized as
 \e\label{cpcommu-rela}
 \left[a_{\lambda_1}(\mathbf{k}_1),a_{\lambda_2}^{\dagger}(\mathbf{k}_2)\right]=\frac{1}{\Mp^2}(2\pi)^3 \delta^3(\mathbf{k}_1-\mathbf{k}_2)\delta_{\lambda_1 \lambda_2}~.
 \q
From Eq.~\eqref{cphsk}, the operator $h^\lambda_{\mathbf{k}}$ satisfies
\m
h^\lambda_{\mathbf{k}} = (h^\lambda_{-\mathbf{k}})^{\dagger}.
\label{cpherm}
\n
On the other hand, since gravity fluctuations $h_{ij}$ are tensor perturbations, $h_{ij}$ follows
\e
{\mathcal{P}}h_{ij}(\eta,\mathbf{x}){\mathcal{P}}^{-1}=h_{ij}(\eta,-\mathbf{x})\ .
\q
under parity transformation, or equivalently
\m
{\mathcal{P}}h^\lambda_{\mathbf{k}}{\mathcal{P}}^{-1}=h^{-\lambda}_{-\mathbf{k}}\ .
\label{cpparity}
\n
Since GR as well as the vacuum state is parity invariant (namely ${\mathcal{P}}|0\rangle = |0\rangle$), the $n$-point function of graviton satisfies
\e
\langle h^{\lambda_1}_{\mathbf{k}_1}\cdots h^{\lambda_n}_{\mathbf{k}_n}\rangle=
\langle {\mathcal{P}}^{-1}{\mathcal{P}}h^{\lambda_1}_{\mathbf{k}_1}{\mathcal{P}}^{-1}
\cdots {\mathcal{P}}h^{\lambda_n}_{\mathbf{k}_n}{\mathcal{P}}^{-1}{\mathcal{P}}\rangle=
\langle{\mathcal{P}}h^{\lambda_1}_{\mathbf{k}_1}{\mathcal{P}}^{-1}
\cdots {\mathcal{P}}h^{\lambda_n}_{\mathbf{k}_n}{\mathcal{P}}^{-1}\rangle \ .
\q
Combining with Eqs.~\eqref{cpherm} and \eqref{cpparity}, we find
\e\label{cp1}
 \langle h^{\lambda_1}_{\mathbf{k}_1}\cdots h^{\lambda_n}_{\mathbf{k}_n}\rangle=
 \langle h^{-\lambda_1}_{-\mathbf{k}_1}\cdots h^{-\lambda_n}_{-\mathbf{k}_n}\rangle=
 \langle h^{-\lambda_1\dagger}_{\mathbf{k}_1}\cdots h^{-\lambda_n\dagger}_{\mathbf{k}_n}\rangle=
 \langle h^{-\lambda_1}_{\mathbf{k}_1}\cdots h^{-\lambda_n}_{\mathbf{k}_n}\rangle^*
 \q
which indicates that the $n$-point function for helicity state possess the symmetry that it becomes its complex conjugation when interchanging $+2 \leftrightarrow -2$.

When all $\mathbf{k}_i$ lie in the same plane, without loss of generality by assuming that all $\mathbf{k}_i$ lie in $y$-$z$ plane, there is another symmetry under the mirror reflection about $y$-$z$ plane.
The transformation matrix is $\mathrm{M}_{yz}=\mathrm{diag}\{-1, 1, 1\}$. Denote $\mathbf{x}=(x,y,z)$, $\mathbf{k}=(k_x,k_y,k_z)$, then $\mathrm{M}_{yz} \mathbf{x}=(-x,y,z)\equiv\tilde{\mathbf{x}}$, $\mathrm{M}_{yz} \mathbf{k}=(-k_x,k_y,k_z)\equiv\tilde{\mathbf{k}}$, and $\mathrm{M}_{yz} e^\lambda(\mathbf{k}) \mathrm{M}_{yz}^{\mathrm{T}}\equiv\tilde{e}^\lambda(\mathbf{k})$. Thus under such a mirror reflection,
\e
\mathcal{M}_{yz} h_{ij}(\eta, \mathbf{x}) \mathcal{M}_{yz}^{-1}= (\mathrm{M}_{yz})_{ik} (\mathrm{M}_{yz})_{jl} h_{kl}(\eta,\tilde{\mathbf{x}})\ ,
\q
or equivalently,
\e
\sum_{\lambda}e^{\lambda}_{ij}(\mathbf{k}) \mathcal{M}_{yz} h^\lambda_{\mathbf{k}} \mathcal{M}_{yz}^{-1}
= \sum_{\lambda} \tilde{e}^{\lambda}_{ij}(\tilde{\mathbf{k}}) h^\lambda_{\tilde{\mathbf{k}}}\ .
\q
Since $\mathbf{k}$ is in the $y$-$z$ plane, $k_x=0$, $\mathbf{k}=\tilde{\mathbf{k}}$, and then the above equation reads
\e\label{eq:Z}
\sum_{\lambda}e^{\lambda}_{ij}(\mathbf{k}) \mathcal{M}_{yz} h^\lambda_{\mathbf{k}} \mathcal{M}_{yz}^{-1}
= \sum_{\lambda} \tilde{e}^{\lambda}_{ij}(\mathbf{k}) h^\lambda_{\mathbf{k}}.
\q
For any $\mathbf{k}$ lies in $y$-$z$ plane, $e^{\lambda}(\mathbf{k})$ can be expressed by
$e^\lambda(\mathbf{k})=R_1(\theta)e^\lambda(\hat{\mathbf{z}})R_1(\theta)^{\mathrm{T}}$, where the definition of
$R_1(\theta)$ is given in \eqref{transmatrix}. Notice that $\mathrm{M}_{yz}$ commutes with $R_1(\theta)$, and hence
\e
\tilde{e}^\lambda(\mathbf{k})=\mathrm{M}_{yz} e^\lambda(\mathbf{k}) \mathrm{M}_{yz}^{\mathrm{T}}
%= \mathrm{M}_{yz} R_1(\theta)e^\lambda(\hat{z})R_1(\theta)^{\mathrm{T}} \mathrm{M}_{yz}^{\mathrm{T}}
=  R_1(\theta) \mathrm{M}_{yz} e^\lambda(\hat{\mathbf{z}}) \mathrm{M}_{yz}^{\mathrm{T}} R_1(\theta)^{\mathrm{T}}\ .
\q
From Eq.~\eqref{CPtensor}, one can easily get
\m
\mathrm{M}_{yz} e^{\lambda}(\mathbf{k}) \mathrm{M}_{yz}^{\mathrm{T}} =  e^{-\lambda}(\mathbf{k}),
\n
and then
\m
\tilde{e}^\lambda(\mathbf{k})=e^{-\lambda} (\mathbf{k}).
\n
Combining with Eq.~\eqref{eq:Z}, we find
\m \label{cpz}
\mathcal{M}_{yz} h^{\lambda}_{\mathbf{k}} \mathcal{M}_{yz}^{-1} = h^{-\lambda}_{\mathbf{k}}.
\n
Therefore,
\e
 \langle h^{\lambda_1}_{\mathbf{k}_1}\cdots h^{\lambda_n}_{\mathbf{k}_n}\rangle=
 \langle h^{-\lambda_1}_{\mathbf{k}_1}\cdots h^{-\lambda_n}_{\mathbf{k}_n}\rangle^*=
 \langle h^{\lambda_1}_{\mathbf{k}_1}\cdots h^{\lambda_n}_{\mathbf{k}_n}\rangle^*
 \q
which indicates that the $n$-point function for circularly polarized states is always real when all momenta are in the same plane. This conclusion is quite nontrivial since $h^{\lambda}_{\mathbf{k}}$ is not a Hermitian operator for the circularly polarized state, i.e.  $h^{\lambda}_{\mathbf{k}}\neq (h^{\lambda}_{\mathbf{k}})^\dagger$ (see Eq.~\eqref{cpherm}).

The bispectrum (three-point function) of graviton is real for the circularly polarized states because the three momentum vectors must lie in the same plane due to the energy-momentum conservation. But the $n(\geq 4)$-point function for the circularly polarized state is complex in general.

\subsection{Linear polarization}

The linearly polarized states can be given by a linear combination of circularly polarized states, namely
\m\label{relation}
h^{+}_{\mathbf{k}} & = & \frac{1}{\sqrt{2}}(h^{+2}_{\mathbf{k}}+h^{-2}_{\mathbf{k}} ), \\
h^{\times}_{\mathbf{k}} & = & \frac{i}{\sqrt{2}}(h^{+2}_{\mathbf{k}}-h^{-2}_{\mathbf{k}} ),
\n
and the linear polarization tensors are related to the circular ones by
\m
e^{+}_{ij}(\mathbf{k}) & = & \frac{1}{\sqrt{2}}(e^{+2}_{ij}(\mathbf{k})+e^{-2}_{ij}(\mathbf{k}))
=\sqrt{2}
\left(
  \begin{array}{ccc}
      1 & 0 & 0\\
    0 & -1 & 0\\
    0 & 0 & 0\\
  \end{array}
  \right),  \\
e^{\times}_{ij}(\mathbf{k}) & = & \frac{-i}{\sqrt{2}}(e^{+2}_{ij}(\mathbf{k})-e^{-2}_{ij}(\mathbf{k}))
=\sqrt{2}
\left(
  \begin{array}{ccc}
    0 & 1 & 0\\
    1 & 0 & 0\\
    0 & 0 & 0\\
  \end{array}
  \right),
\n
where $+$ and $\times$ denote two linearly polarized states.
One can easily find that the linear polarization tensors satisfy
\m
e_{ij}^{+}(\mathbf{k})=e_{ij}^{+}(-\mathbf{k}),\quad e_{ij}^{\times}(\mathbf{k})=-e_{ij}^{\times}(-\mathbf{k}).
\n
For linear polarization, the most general solution of quantized tensor perturbation becomes
 \m
 h^{s}_{\mathbf{k}}(\eta)&\equiv& \psi_k a_s(\mathbf{k})+(-1)^{1-s\over 2} \psi^*_k a_s^\dagger(-\mathbf{k}),
 \label{lphsk}
 \n
where $s=+1,\ -1$ for $+,\ \times$ polarized states respectively. Here $\psi_k$ is same as that in circular polarization, and the commutators between the creation and annihilation operators are
\e\label{lpcommu-rela}
 \left[a_{s_1}(\mathbf{k}_1),a_{s_2}^{\dagger}(\mathbf{k}_2)\right]=\frac{1}{\Mp^2}(2\pi)^3 \delta^3(\mathbf{k}_1-\mathbf{k}_2)\delta_{s_1 s_2}~.
 \q
From Eq.~\eqref{lphsk}, the operator $h^s_{\mathbf{k}}$ satisfies
\m
h^+_{\mathbf{k}} = (h^+_{-\mathbf{k}})^{\dagger}, \quad h^\times_{\mathbf{k}} = -(h^{\times}_{-\mathbf{k}})^{\dagger}.
\label{lpherm}
\n
Similar to the circular polarization, under the parity transformation the operator $h^s_{\mathbf{k}}$ follows
\m
{\mathcal{P}}h^+_{\mathbf{k}}{\mathcal{P}}^{-1}=h^{+}_{-\mathbf{k}}, \quad
{\mathcal{P}}h^\times_{\mathbf{k}}{\mathcal{P}}^{-1}= -h^{\times}_{-\mathbf{k}} \ .
\label{lpparity}
\n
Therefore
\e
 \langle h^{s_1}_{\mathbf{k}_1}\cdots h^{s_n}_{\mathbf{k}_n}\rangle
 %= (-1)^l \langle h^{s_1}_{-\mathbf{k}_1}\cdots h^{s_n}_{-\mathbf{k}_n}\rangle= (-1)^l (-1)^l\langle h^{s_1\dagger}_{\mathbf{k}_1}\cdots h^{s_n\dagger}_{\mathbf{k}_n}\rangle
= \langle h^{s_1}_{\mathbf{k}_1}\cdots h^{s_n}_{\mathbf{k}_n}\rangle^*,
\label{lpnpf}
 \q
where Eqs.~\eqref{lpherm} and \eqref{lpparity} are taken into account. We conclude that the $n$-point functions of gravitational waves in terms of linear polarized state are always real whether all $\mathbf{k}_i$ lie in the same plane or not.
Because a real $n$-point function is convenient in analysis and the $n$-point function in terms of circularly polarized state can be easily derived from $n$-point function in terms of linearly polarized state by using \eqref{relation}, we will compute the four-point correlation function of graviton in terms of linearly polarized state in this paper.

Before closing this subsection, we also want to illustrate some simple properties of $n$-point function of graviton when all $\mathbf{k}_i$ are in the same plane. Similar to the former subsection, assuming that all $\mathbf{k}_i$ lie in the $y$-$z$ plane, and under a mirror reflection $M_{yz}$ transformation, we have
\m
\mathrm{M}_{yz} e^{+}(\mathbf{k}) \mathrm{M}_{yz}^{\mathrm{T}} = e^{+}(\mathbf{k}), \quad
\mathrm{M}_{yz} e^{\times}(\mathbf{k}) \mathrm{M}_{yz}^{\mathrm{T}} =-e^{\times}(\mathbf{k}),
\n
and
\m
\mathcal{M}_{yz} h^{+}_{\mathbf{k}} \mathcal{M}_{yz}^{-1} =  h^{+}_{\mathbf{k}}, \quad
\mathcal{M}_{yz} h^{\times}_{\mathbf{k}} \mathcal{M}_{yz}^{-1} =-h^{\times}_{\mathbf{k}}.
\n
Denoting $m$ as the number of $\times$ linear polarized modes in the $n$-point function, we have
\e
\langle h^{s_1}_{\mathbf{k}_1}\cdots h^{s_n}_{\mathbf{k}_n}\rangle =
(-1)^m \langle h^{s_1}_{\mathbf{k}_1}\cdots h^{s_n}_{\mathbf{k}_n}\rangle\ .
\q
Thus the $n$-point function for linearly polarized state equals zero if $m$ is an odd number when all the momentum vectors lie in the same plan. This property can be seen explicitly for the four-point function in the next section.

\section{The four-point correlation function of graviton in de Sitter background}
\label{section:4pt}

In this section we will use the in-in formalism to calculate the four-point correlation function of graviton during a period of exactly de Sitter expansion, including the contact and exchange diagrams. A brief review of tensor power spectrum and the three-point function of graviton is given in Appendix \ref{app:3pt}.

\subsection{In-in formalism}

In this subsection we will introduce the Schwinger-Keldysh \emph{in-in} formalism \cite{Weinberg:2005vy} to calculate the $n$-point function of gravity fluctuations during inflation. The expectation value of some product $Q(t)$ of field operators at fixed time $t$ is given by
 \e
 \langle Q(t)\rangle=\langle in|Q(t)|in\rangle,
 \q
where $|in\rangle$ is the initial state at very early times when the wavelength is deep inside the horizon. In the interaction picture, we have
\e\label{eq:in-in}
  \langle Q(t)\rangle =
  {\langle 0|
    \left[  \bar T e^{i \int_{t_0}^t H_I(\tau')d\tau'}\right]  Q^I(t)
    \left[  T e^{-i \int_{t_0}^t H_I(\tau'')d\tau''}\right]
  |0\rangle} ~,
\q
where $H_I$ is the interaction part of the Hamiltonian. Here $T$ denotes a time-ordered product, $\bar T$ denotes an anti-time-ordered product, and $Q^I$ is the product $Q$ in the interaction picture. Utilizzing the Dyson series, one can get
 \begin{align}\label{eq:in-in-comm}
  \langle Q(t)\rangle = \sum_{n=0}^\infty  i^n
  \int_{t_0}^{t} dt_1 \int_{t_0}^{t_1} dt_2 \cdots \int_{t_0}^{t_{n-1}} dt_n
  \left\langle [H_I(t_n), [H_I(t_{n-1}), \cdots,
  [H_I(t_1),Q^I(t)]\cdots]]\right\rangle~,
\end{align}
which is called the commutator form. Note that the $n$-th order in-in formalism have $n$ commutators. Supposing that operator $Q$ is Hermitian, i.e. $Q=Q^\dagger$, the first and second order terms read
\begin{align}\label{eq:in-in-1}
  \langle Q(t) \rangle_1 = 2\mathrm{Im} \int_{t_0}^{t} dt_1 \langle0| Q^I(t) H_I(t_1) |0\rangle~,
\end{align}
\m
  \langle Q(t) \rangle_2 &=& \int_{t_0}^t dt_1 \int_{t_0}^t dt_2 \langle0| H_I(t_1) Q^I(t) H_I(t_2) |0\rangle \nonumber \\
 &-& 2\mathrm{Re} \int_{t_0}^t dt_1 \int_{t_0}^{t_1}dt_2 \langle0| Q^I(t) H_I(t_1) H_I(t_2) |0\rangle ~.
 \label{eq:in-in-2}
\n
Here we need to stress that we should use Eq.~\eqref{eq:in-in-comm} to calculate the $n$-point function if  the operator $Q$ is not Hermitian.
Since there are no time derivative terms in the cubic order interaction Lagrangian for the gravity fluctuations in de Sitter background, the interaction part of Hamiltonian follows that $H_I=-\int dx^3 {\cal L}_I$ up to the fourth order.

In general the the $n$-point correlation function of gravitational waves in the de Sitter background  takes the form
\e
\langle h^{s_1}_{\mathbf{k}_1}h^{s_2}_{\mathbf{k}_2}\cdots h^{s_n}_{\mathbf{k}_n} \rangle =
(2\pi)^3 \delta^{(3)}(\sum_{i=1}^n \mathbf{k}_i) \left(\frac{H}{\Mp} \right)^{2(n-1)}
{\mathcal{T}^{s_1 s_2 \cdots s_n}(\mathbf{k}_1,\mathbf{k}_2,\cdots,\mathbf{k}_n)}\ ,
\label{Tnp}
\q
and
\e
{\mathcal{T}^{s_1 s_2 \cdots s_n}(\mathbf{k}_1,\mathbf{k}_2,\cdots,\mathbf{k}_n)} =
\frac{{\mathcal{A}}^{s_1 s_2 \cdots s_n}(\mathbf{k}_1,\mathbf{k}_2,\cdots,\mathbf{k}_n)}{\prod_{i=1}^n k_i^3}\ ,
\q
where $\delta^{(3)}(\sum_{i=1}^n \mathbf{k}_i)$ implies the energy-momentum conservation. Furthermore, the constraints on the function of ${\mathcal{T}^{s_1 s_2 \cdots s_n}(\mathbf{k}_1,\mathbf{k}_2,\cdots,\mathbf{k}_n)}$ from the de Sitter isometries are discussed in Appendix \ref{app:isometries}. In the next two subsections we will use the in-in formalism to explicitly calculate the amplitude ${\mathcal{A}}^{s_1 s_2 s_3 s_4}(\mathbf{k}_1,\mathbf{k}_2,\mathbf{k}_3,\mathbf{k}_4)$ which is expected to be different for different interaction term.

\subsection{Contact diagram}
\label{section:Contact}

\begin{figure*}[htbp]
\centering
\begin{center}
\includegraphics[scale=0.5]{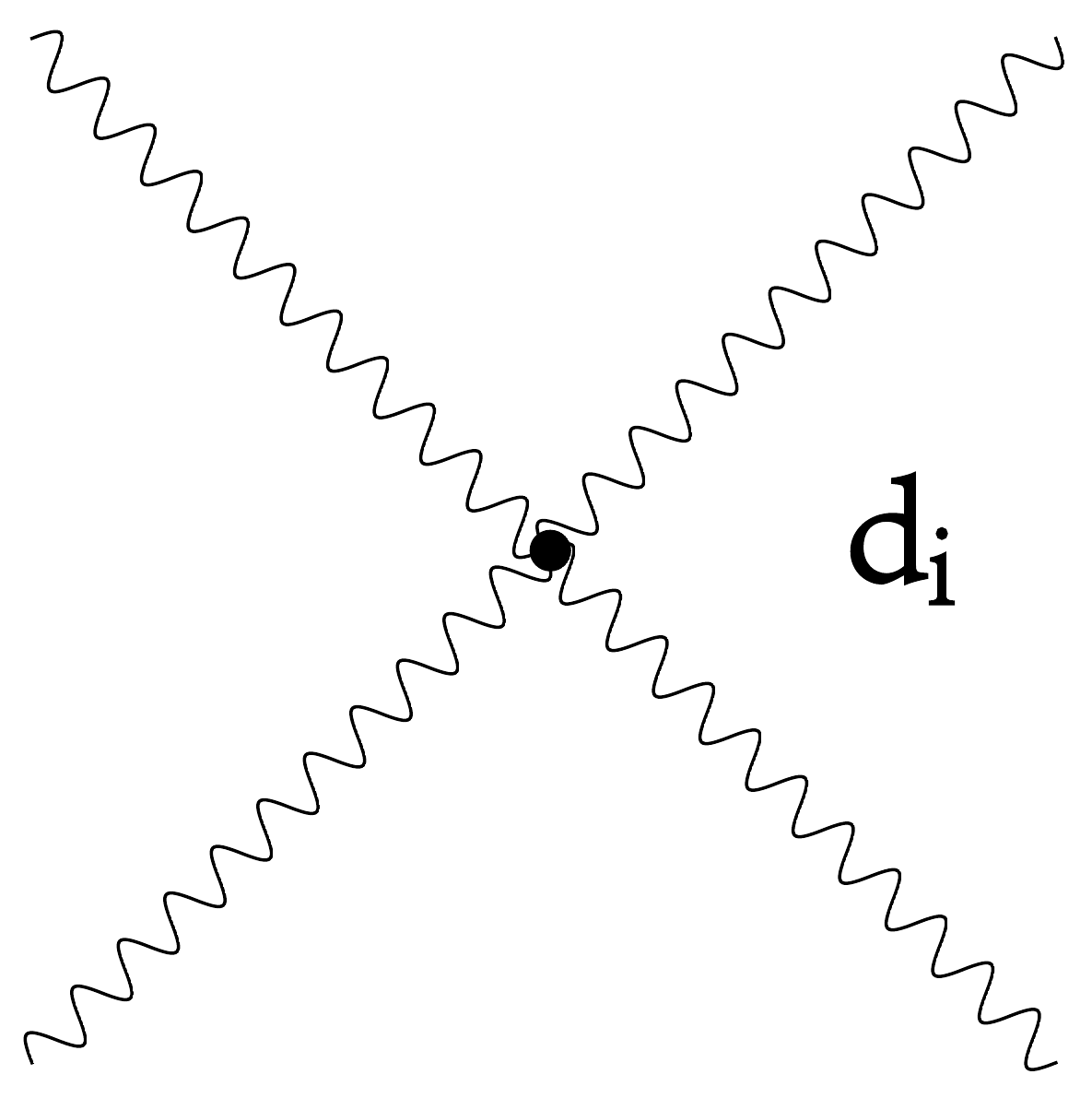}
\end{center}
\caption{Contact diagram. }\label{fig:contactdiagram}
\end{figure*}

In this subsection we will work out the contribution to the four-point correlation function of graviton from the contact diagram (see Fig.~\ref{fig:contactdiagram}).
At the tree level, we only need to use the first order in-in formalism
\begin{align}
  \langle h^{s_1}_{\mathbf{k}_1}h^{s_2}_{\mathbf{k}_2}h^{s_3}_{\mathbf{k}_3}h^{s_4}_{\mathbf{k}_4}\rangle_c =  i  \int_{-\infty}^{0} d\eta_1 \langle0| \[H^{(4)}_I(\eta_1) ,h^{s_1}_{\mathbf{k}_1}h^{s_2}_{\mathbf{k}_2}h^{s_3}_{\mathbf{k}_3}h^{s_4}_{\mathbf{k}_4}\] |0\rangle~,
  \label{in-in:contact}
\end{align}
where $H^{(4)}_I=-\int d^3x\mathcal{L}^{(4)}_I $. Remind that Eq.~\eqref{eq:in-in-1} is not applicable in this case because the operator $h^{s_1}_{\mathbf{k}_1}h^{s_2}_{\mathbf{k}_2}h^{s_3}_{\mathbf{k}_3}h^{s_4}_{\mathbf{k}_4}$ is not Hermitian. Before using the above in-in formalism to compute four-point function of graviton, we need to find the solution of $\psi^{(2)}$ from the constraint equation \eqref{constraint2}. Considering Eq.~\eqref{dech}, we can easily find
\e
 \psi^{(2)}(\mathbf{x},\eta)=\int \frac{d^3 k_1 d^3 k_2}{(2\pi)^6}\(\tilde{h}^\prime_{ij}(\mathbf{k}_1,\eta)\tilde{h}^\prime_{ij}(\mathbf{k}_2,\eta)-\mathbf{k}_1\cdot
 \mathbf{k}_2 \tilde{h}_{ij}(\mathbf{k}_1,\eta)\tilde{h}_{ij}(\mathbf{k}_2,\eta)\)\frac{e^{i\mathbf{k}_{12}\cdot\mathbf{x}}}{16Hk_{12}^2},
 \label{solpsi2}
 \q
 where $\mathbf{k}_{ij}=\mathbf{k}_i+\mathbf{k}_j $, $k_{ij}^a=k_i^a+k_j^a$ and $k_{ij}=|\mathbf{k}_{ij}|$. Now adopting the in-in formalism in Eq.~\eqref{in-in:contact}, after a lengthy but straightforward computation, we get
\begin{dmath}
{\mathcal{A}}^{s_1 s_2 s_3 s_4}(\mathbf{k}_1,\mathbf{k}_2,\mathbf{k}_3,\mathbf{k}_4)={1\over 16} \bigg\{
-\frac{d_1}{4} \mathcal{I}_1 \frac{\mathbf{k}_{12}\cdot\mathbf{k}_4}{16 k_{12}^2}e^1_{ij}e^2_{ij}e^3_{kl}e^4_{kl}
+ \frac{d_2}{48} \mathcal{I}_2 (k_1^2 k_2^2- k_1^2 k_3^2 ) e^1_{ij}e^2_{jk}e^3_{kl}e^4_{li}+
\mathcal{I}_3
\left[\frac{d_3}{48}( \mathbf{k}_1 \cdot \mathbf{k}_3 - \mathbf{k}_1 \cdot \mathbf{k}_2 )
 e^1_{ij}e^2_{jk}e^3_{kl}e^4_{li}
-\frac{d_4}{16} e^1_{ij}e^2_{ij}k^l_{1}e^3_{lm}e^4_{mn}k^n_{2} +\frac{d_5}{8}\(-\frac{1}{2} k^i_2 k^m_3+ k_3^i k_2^m \)e^1_{ij}e^2_{jk}e^3_{kl}e^4_{lm} \right]
 \bigg\}
   + \text{ 23 perms.},
\end{dmath}
where
\begin{eqnarray}
\mathcal{I}_1 &= & {1\over K^4}
\big\{
-4(K+3k_4)k_1^2 k_2^2 k_3^2 - 2 \mathbf{k}_1\cdot\mathbf{k}_2k_3^2[6 k_1 k_2 k_4+K^2(2K-k_3)
\nonumber\\
&&+2 K(k_1 k_2 +k_1 k_4 +k_2 k_4) ]
\big\}, \\
\mathcal{I}_2 &= & \frac{1}{K^3}\left[K\left(K+k_3+k_4\right)+2k_3 k_4 \right],
\\
\mathcal{I}_3 &= & 2\bigg\{ K -\frac{1}{K^3}\big[2k_1 k_2 k_3 k_4+K^2(k_1 k_2 +k_1 k_3+ k_1 k_4
+k_2 k_3+ k_2 k_4 +k_3 k_4)
\nonumber\\
&&+K(k_1 k_2 k_3 +k_1 k_3 k_4 +k_1 k_2 k_4 + k_2 k_3 k_4)\big)
 \bigg\},
\end{eqnarray}
$K\equiv k_1+k_2+k_3+k_4$, and $e^l_{ij}\equiv e^{s_l}_{ij}(\mathbf{k}_l)$. Here the divergent parts from performing the time integration are counteracted. In order to sketch out some simple features of the four-point function of graviton from the contact diagram, we will consider some special configurations in the momentum space.

\begin{figure*}[htbp]
\centering
\begin{center}
\includegraphics[scale=1]{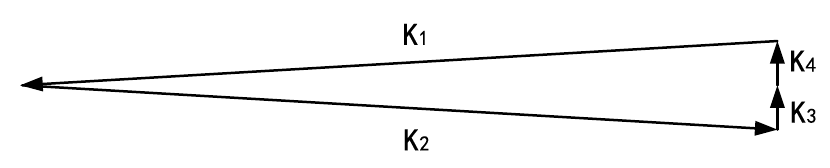}
\end{center}
\caption{Bilateral squeezed limit.}\label{bilateral}
\end{figure*}

First of all, we consider the bilateral squeezed limit in which $\mathbf{k}_3=\mathbf{k}_4=-\mathbf{k}_{12}/2$, $k_1=k_2=\tilde{k}$ and $k_3=k_4=k\rightarrow 0$. See Fig.~\ref{bilateral}.
In this limit, all $\mathbf{k}_i$ lie in the same plane, and there are only four non-zero amplitude, namely
\m
\mathcal{T}^{\times\times\times\times}&=&-\frac{1}{96\tilde{k}^3k^6}(d_2+6d_3+36d_4-18d_5)=-\frac{25}{96}{1\over \tilde{k}^3k^6}, \\
\mathcal{T}^{\times\times++}&=&-\frac{1}{96\tilde{k}^3k^6}(d_2+6d_3+36d_4)=-\frac{43}{96}{1\over \tilde{k}^3k^6}, \\
\mathcal{T}^{++\times\times}&=&\frac{1}{96\tilde{k}^3k^6}(d_2+6d_3+36d_4-18d_5)=\frac{25}{96}{1\over \tilde{k}^3k^6}, \\
\mathcal{T}^{++++}&=&-\frac{1}{192\tilde{k}^3k^6}(d_2-72d_4)=\frac{71}{192}{1\over \tilde{k}^3k^6}.
\n
We see that the interaction term denoted by $d_1$ in the fourth order action does not make a contribution to the four-point function in the bilateral squeezed limit, and all of the non-zero amplitudes have the same scaling behavior $\langle hhhh\rangle\propto {\mathcal{T}}\propto 1/(\tilde{k}^3 k^6)$ which is divergent as $1/k^6$ in the limit of $k\rightarrow 0$. This is similar to the local-form four-point function of scalar perturbation from the contact diagram.

%Consequently, in this limit, the 4-point function $\langle hhhh\rangle\propto {\mathcal{T}}\propto 1/(\tilde{k}^3 k^6) $. At first glance, it seems weird that the trispectrum doesn't own the symmetry of switching $\times \leftrightarrow +$ state, because of GR's parity invariant.  This argument, however, is specious in that helicity state is the eigenstate of angular momentum while linear polarized state is not. As we have proved Eq \eqref{cpparity} \eqref{lpparity}, if we work in helicity state, the 4-point function possess a symmetry of interchanging $+2 \leftrightarrow -2$, while linear polarized state has nothing to do with this property. There is another paradox that ${\mathcal{T}}$ would have owned the permutation symmetry of $s_1,s_2,s_3,s_4$. And this puzzle can be settled by noticing ${\mathcal{T}}^{s_1 s_2 s_3 s_4}(\mathbf{k}_1,\mathbf{k}_2,\mathbf{k}_3,\mathbf{k}_4)={\mathcal{T}}^{s_2 s_1 s_3 s_4}(\mathbf{k}_2,\mathbf{k}_1,\mathbf{k}_3,\mathbf{k}_4)\neq {\mathcal{T}}^{s_2 s_1 s_3 s_4}(\mathbf{k}_1,\mathbf{k}_2,\mathbf{k}_3,\mathbf{k}_4)$. Lastly, as proven above, in bilateral squeezed limit all momentums are in the same plane, trispectrum equals zero if there are odd $\times$ modes.

\begin{figure*}[htbp]
\centering
\begin{center}
$\begin{array}{c@{\hspace{0.2in}}c} \multicolumn{1}{l}{\mbox{}} &
\multicolumn{1}{l}{\mbox{}}\\
\includegraphics[scale=0.5]{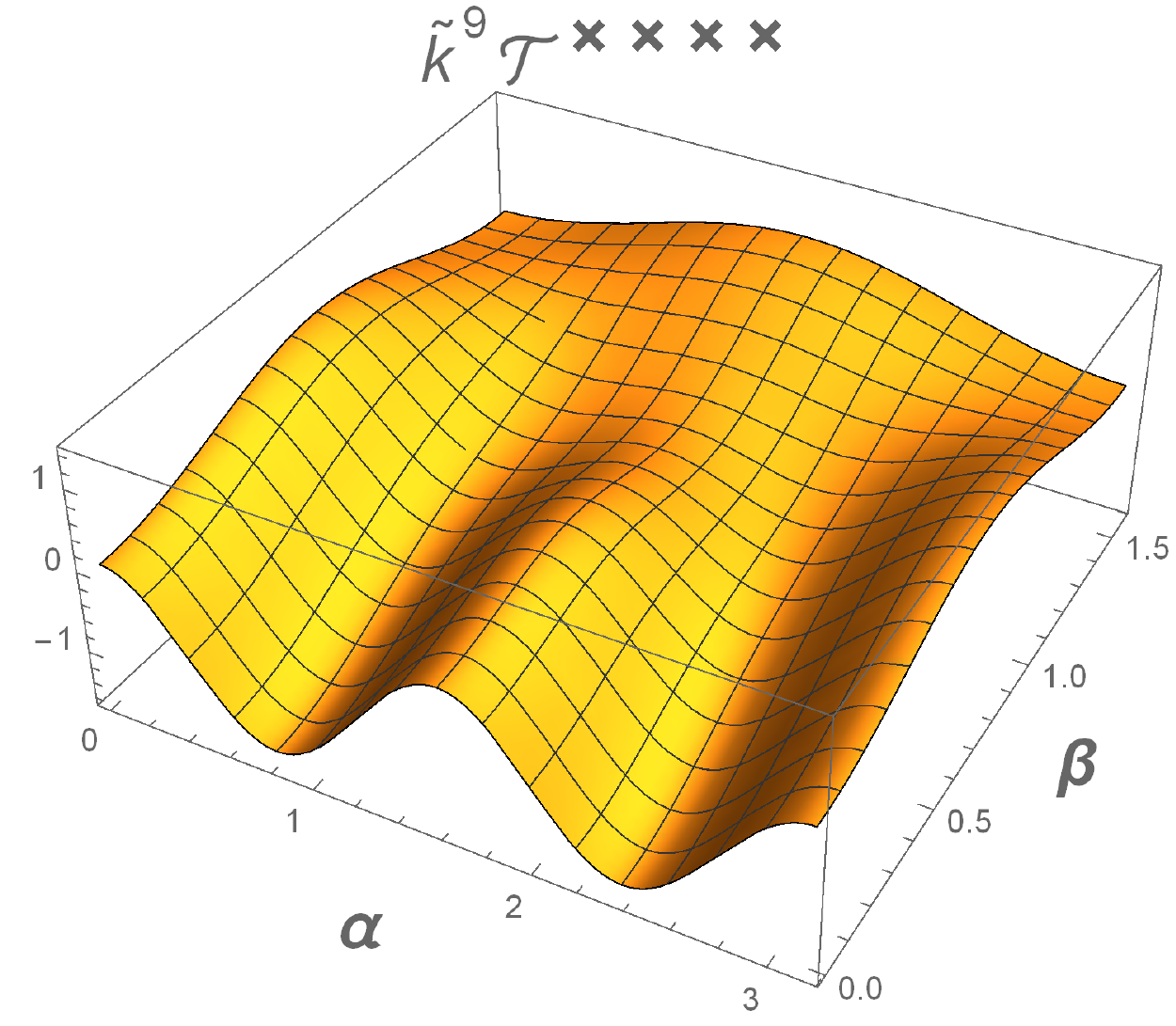} &\includegraphics[scale=0.5]{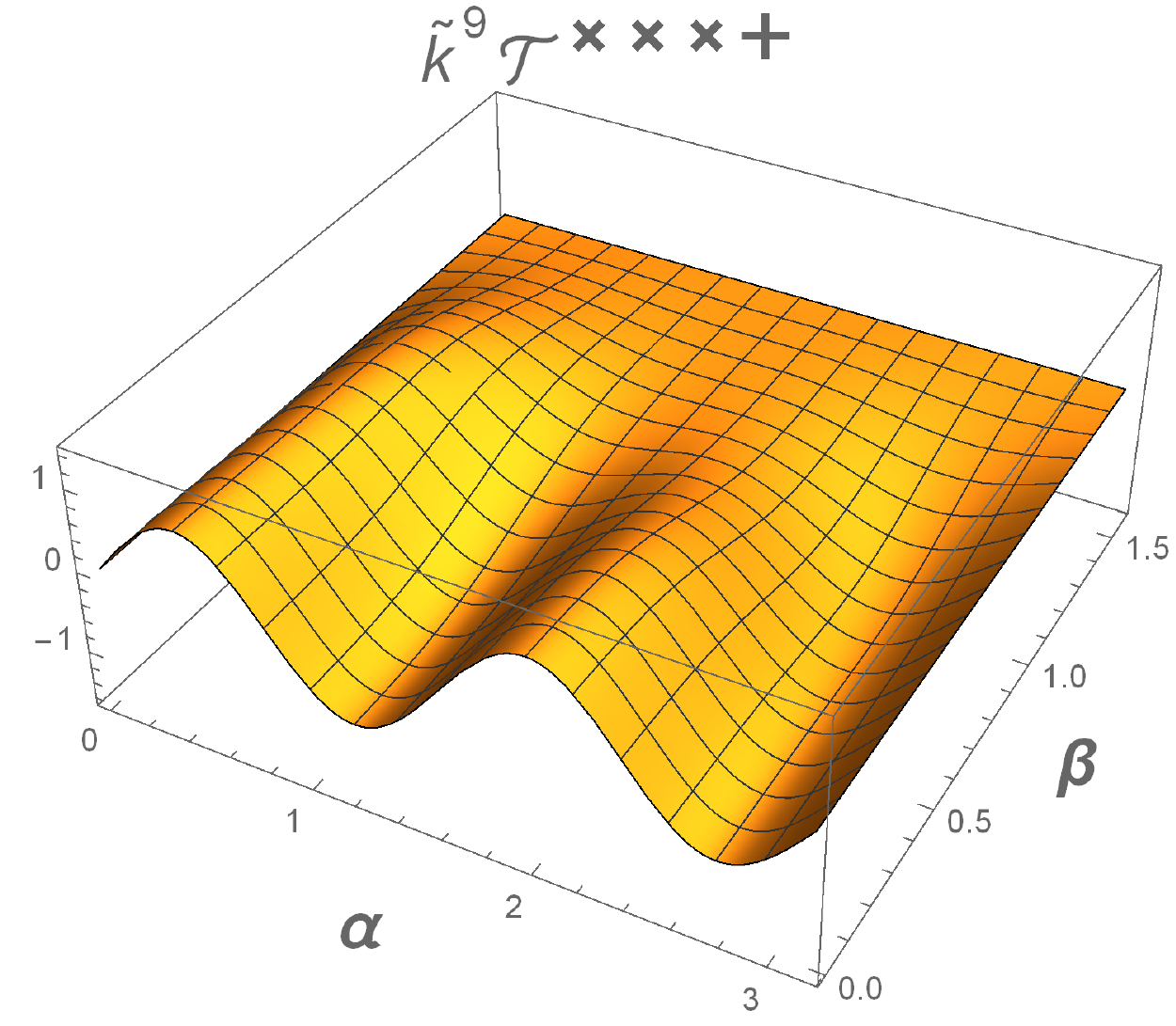}\\
\includegraphics[scale=0.5]{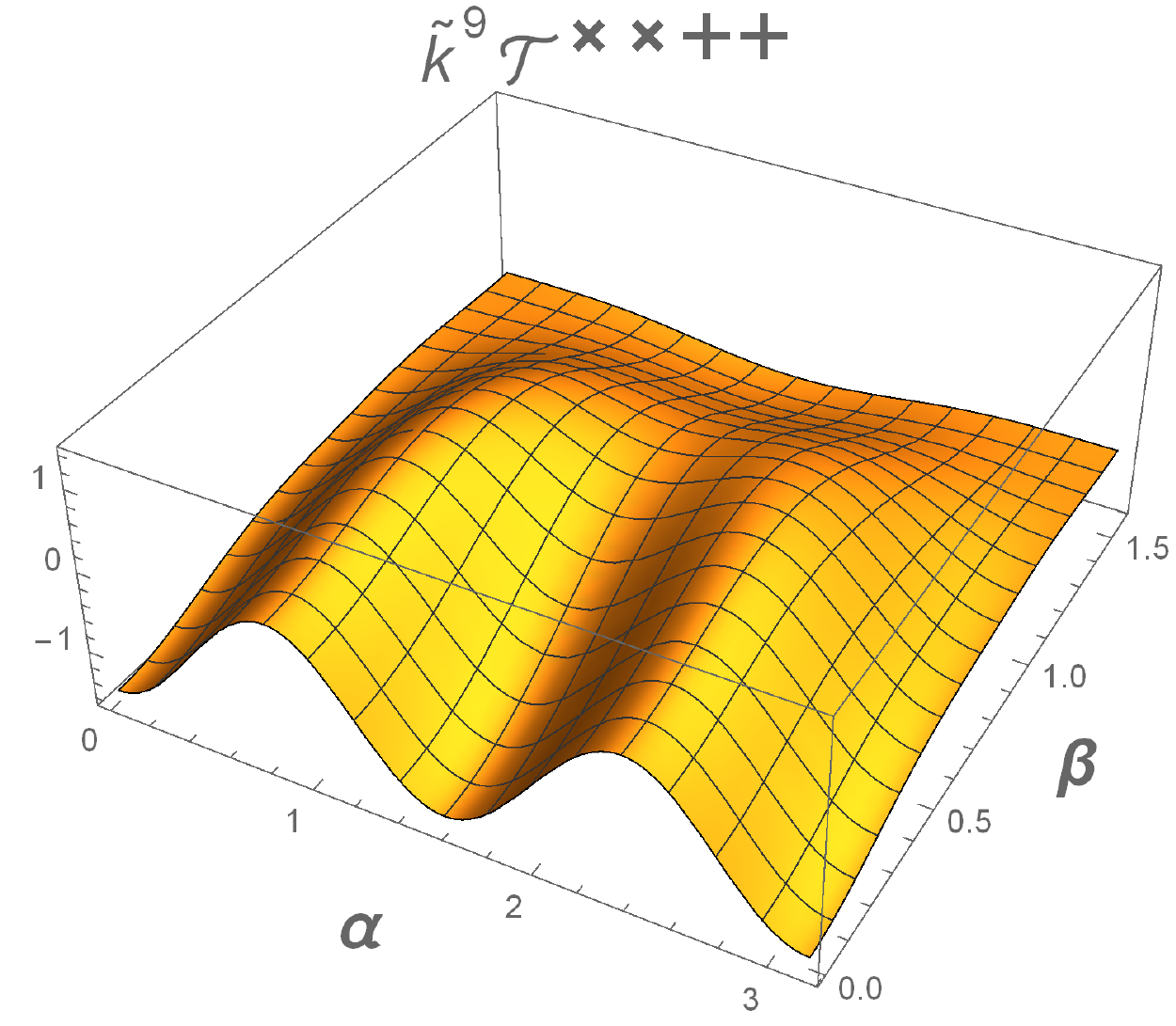} &\includegraphics[scale=0.5]{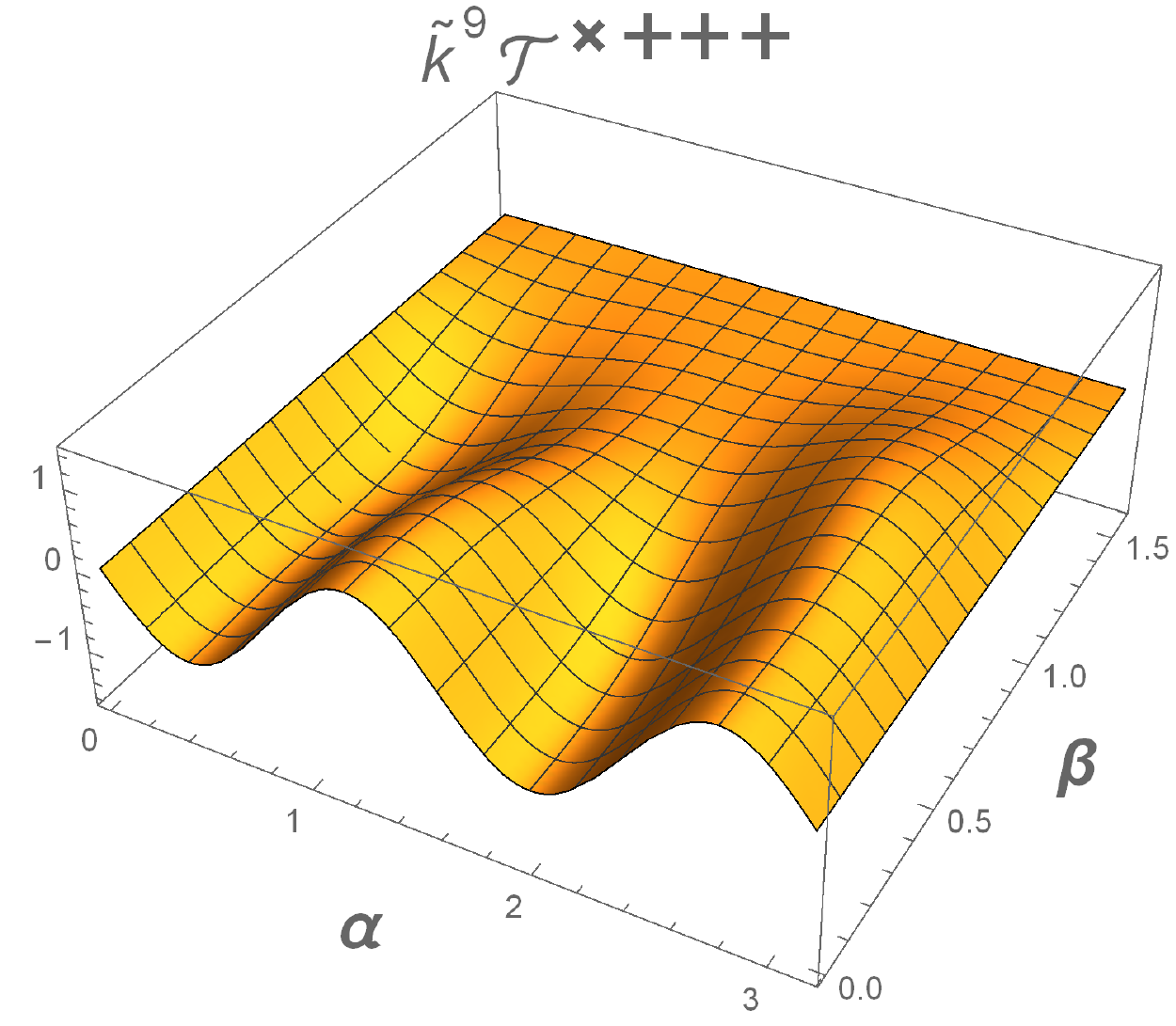}\\
\includegraphics[scale=0.5]{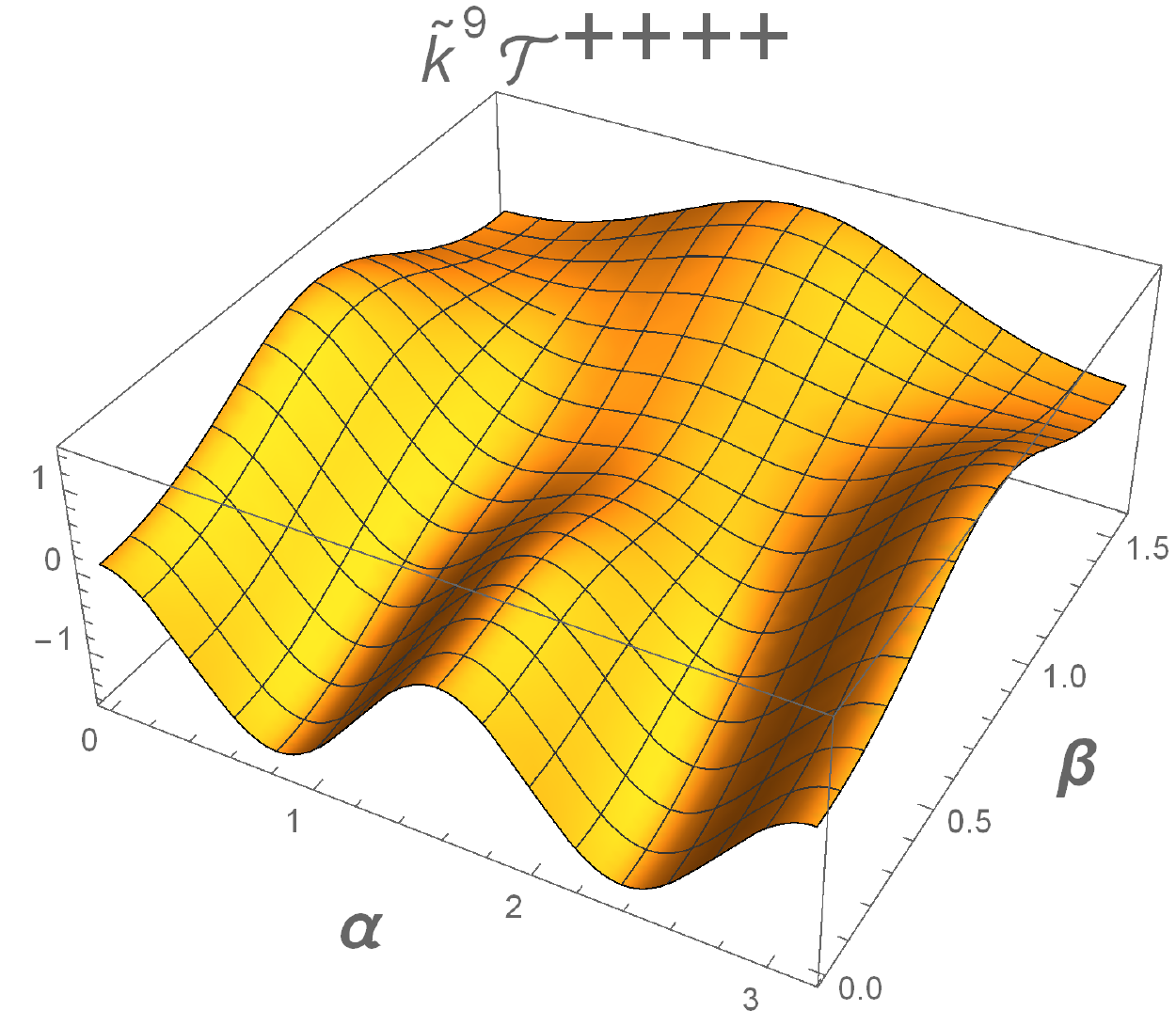} &\\
%\mbox{(a)} & \mbox{(b)}
\end{array}$
\end{center}
\caption{The five independent shapes of trispectrum from contact diagram in the equilateral quadrangle momentum configuration.
%In this group of figures, we consider the equilateral quadrangle, which is $k_1=k_2=k_3=k_4=\tilde{k}$ and $\theta_i=\phi_i=\beta$, and plot all different kinds of $\tilde{k}^9{\mathcal{T}}$ of contact diagram with respect to $\alpha$ and $\beta$.
}
\label{fig:contact}
\end{figure*}

\begin{figure*}[htbp]
\centering
\begin{center}
\includegraphics[scale=1]{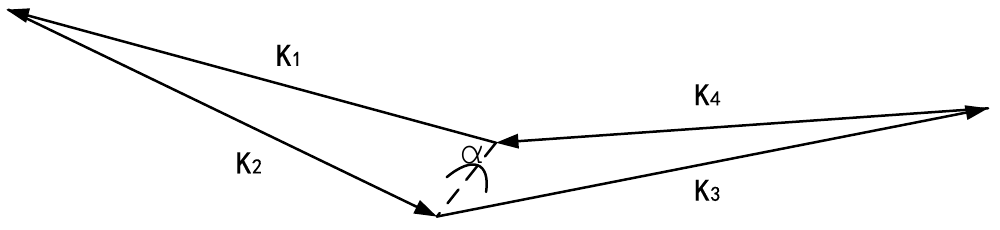}
\end{center}
\caption{Diamond squeezed limit.}\label{diamond}
\end{figure*}

Secondly, we consider the equilateral quadrangle momentum configuration in which $k_1=k_2=k_3=k_4=\tilde{k}$ and $\theta_1=\theta_2=\phi_1=\phi_2\equiv \beta$. In this configuration, the two angles in Fig.~\ref{figpolarization}, namely $\alpha\in [0, \pi]$ and $\beta\in[0,\pi/2]$, are free parameters. Since there is not a special momentum mode among these four momentum modes, there are only five independent shapes, i.e. $\mathcal{T}^{\times\times\times\times}$, $\mathcal{T}^{\times\times\times+}$, $\mathcal{T}^{\times\times++}$, $\mathcal{T}^{\times+++}$ and $\mathcal{T}^{++++}$. The shapes of these five amplitudes show up in Fig.~\ref{fig:contact}.
$\mathcal{T}^{\times\times\times+}=\mathcal{T}^{\times+++}=0$ for $\alpha=0$ or $\alpha=\pi$ because in these two cases all of four momenta lie in the same plane and there are odd number of $\times$ linearly polarized states.
Furthermore, in the limit of $\beta\rightarrow \pi/2$ corresponding to the diamond squeezed limit (see Fig.\ref{diamond})  in which $k_{12}\rightarrow 0$, the four-point function of graviton from contact diagram is regular.

\subsection{Exchange diagram}

\begin{figure*}[htbp]
\centering
\begin{center}
\includegraphics[scale=0.55]{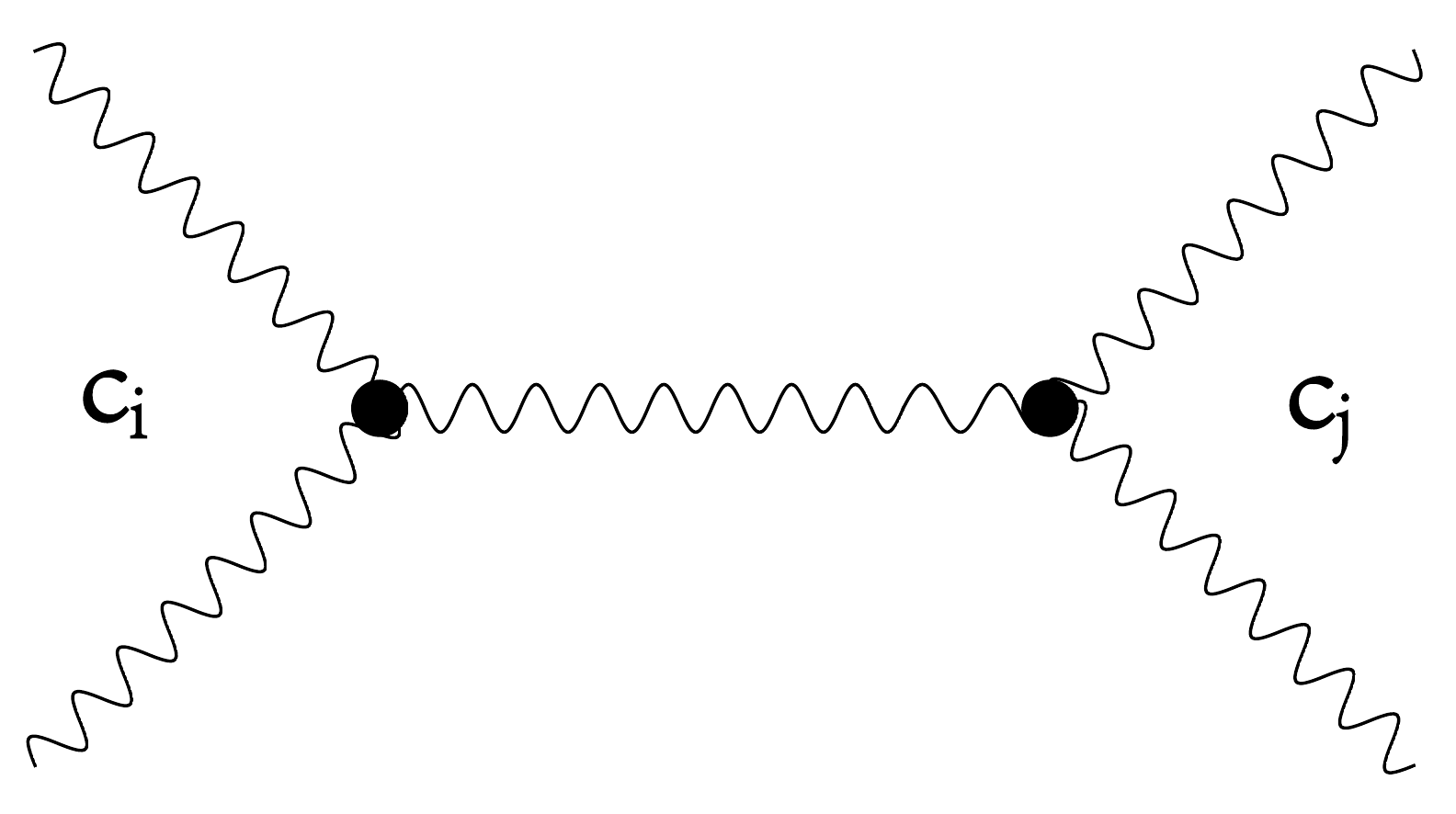}
\end{center}
\caption{Exchange diagram. }\label{fig:exchangediagram}
\end{figure*}

The third order perturbation Hamiltonian can also make a contribution to the four-point function of graviton through the exchange diagram (see Fig.~\ref{fig:exchangediagram}).
Applying the second order in-in formalism, the four-point function from the exchange diagram takes the form
\e
 \label{exch-in-in}
 \langle h^{s_1}_{\mathbf{k}_1}h^{s_2}_{\mathbf{k}_2}h^{s_3}_{\mathbf{k}_3}h^{s_4}_{\mathbf{k}_4} \rangle_e
 =  - \int_{-\infty}^0 d\eta_1 \int_{-\infty}^{\eta_1} d\eta_2  \langle0|\[ H^{(3)}_I(\eta_2) , \[ H^{(3)}_I(\eta_1), h^{s_1}_{\mathbf{k}_1}h^{s_2}_{\mathbf{k}_2}h^{s_3}_{\mathbf{k}_3}h^{s_4}_{\mathbf{k}_4}\]\]   |0\rangle ~,
\q
where the third order interaction Hamiltonian is given by
\m
H^{(3)}_I(\eta) =-\frac{\Mp^2}{4} \int d x^3  a^2(c_1 h_{ik}h_{jl}-\frac{1}{2}c_2 h_{ij} h_{kl})\partial_k \partial_l h_{ij}.
\n
It is remarkable that when doing time double integrals from $-\infty$ to $0$ we will get several divergent terms due to high negative power of conformal time.Fortunately, these divergences  cancel each other exactly when we do permutations of $\mathbf{k}_1$, $\mathbf{k}_2$, $\mathbf{k}_3$ and $\mathbf{k}_4$.
Finally we get the amplitude of the four-point function from the exchange diagram as follows
 \m
&& \mathcal{A}^{s_1 s_2 s_3 s_4}(\mathbf{k}_1,\mathbf{k}_2,\mathbf{k}_3,\mathbf{k}_4) \\
&&= \frac{\mathcal{I}_{1234}}{512k_{12}^3}
 \sum_{s}(-1)^{\frac{s-1}{2}}(c_1 A^s(1,2)-\frac{1}{2}c_2 B^s(1,2))(c_1 A^s(3,4)-\frac{1}{2}c_2 B^s(3,4))
  + \text{ 23 perms}, \nn
 \n
where
\m
\mathcal{I}_{1234}&=& \mathcal{B}_1+\mathcal{B}_2-2\mathcal{B}_3\mathcal{B}_4-2\mathcal{B}_5, \\
\mathcal{B}_1 &=& \((k_{12}+k_1+k_2)-\frac{k_{12}k_1k_2+(k_{12}+k_1+k_2)(k_{12}k_1+k_{12}k_2+k_1 k_2)}{(k_{12}+k_1+k_2)^2}\)\(1\leftrightarrow3,2\leftrightarrow4 \),  \nn\\
\mathcal{B}_2 &=& -k^2_{12}-\frac{1}{2}(k_1^2+k_2^2+k_3^2+k_4^2)-2\(\frac{k_{12} k_1 k_2}{K}
-(k_{12}-k_1 -k_2)K-\frac{3}{4}K^2\),    \nn\\
\mathcal{B}_3 &=& K+\frac{1}{K^2}\(k_1 k_2 k_{12}+K(k_{12}(k_1+k_2)-k_1 k_2)\), \nn\\
\mathcal{B}_4 &=& \frac{k_3 k_4 k_{12}+(k_3+k_4+k_{12})(k_3 k_4+k_{12}(k_3+k_4))}{(k_{12}+k_3+k_4)^2}, \nn\\
\mathcal{B}_5 &=& \frac{k_3k_4k_{12}}{K^3(k_3+k_4+k_{12})}\big\{ K^2(k_{12}-k_1-k_2)+K(k_{12}(k_1+k_2)-2k_1 k_2)+2k_{12} k_1 k_2 \big\}, \nn
\n
and
\m
A^s(i,j)&=&k_{ij}^a e^{s_i}_{ab}(\mathbf{k}_i)e^s_{bc}(\mathbf{k}_{ij})
e^{s_j}_{cd}(\mathbf{k}_j)k_{ij}^d+2k^a_je^s_{ab}(\mathbf{k}_{ij})
e^{s_j}_{bc}(\mathbf{k}_j)e^{s_i}_{cd}(\mathbf{k}_i)k_j^d~, \nn\\
B^s(i,j)&=& k_{ij}^a e^{s_j}_{ab}(\mathbf{k}_j) k_{ij}^b e^{s_i}_{cd}(\mathbf{k}_i)e^s_{cd}(\mathbf{k}_{ij})
+k_j^a e^{s_i}_{ab}(\mathbf{k}_i)k_j^b e^s_{cd}(\mathbf{k}_{ij})e^{s_j}_{cd}(\mathbf{k}_j)\nn\\
&&+k_j^ae^s_{ab}(\mathbf{k}_{ij})k_j^b e^{s_i}_{cd}(\mathbf{k}_i)e^{s_j}_{cd}(\mathbf{k}_j)~.
\n
 Note that $i,j= 1,2,3,4$ represent
$i$th, $j$th momentum and need not do summation, while the indices $a,b,c,d = 1,2,3$ are the component indices.

Analogous to the analysis of contact diagram, in the bilateral squeezed limit, we find
\m
{\mathcal{T}}^{s_1 s_2 s_3 s_4} =
\frac{15}{128}c^2_2(s_1+s_2)(1+s_3)(1+s_4)\frac{1}{\tilde{k}^3 k^6}~,
\n
which indicates that there are only two non-zero components, i.e. ${\mathcal{T}}^{++++}$ and ${\mathcal{T}}^{\times\times++}$ and only the interaction term denoted by $c_2$ in the third order action contributes to the four-point function in the bilateral squeezed limit.
The non-zero components diverge on behavior of $1/k^6$ in the limit of $k\rightarrow 0$ as well.

Secondly, let's consider the diamond squeezed limit, namely $k_1=k_2=k_3=k_4=\tilde k$ and
$k_{12}\rightarrow 0$. In this limit, we have
\m
{\mathcal{T}}^{s_1 s_2 s_3 s_4}=
\frac{9}{64}c_2^2\cos2\alpha(s_1+s_2)(s_3+s_4)\frac{1}{{\tilde k}^6 k_{12}^3}~.
\n
Again we see that only the interaction term denoted by $c_2$ in the third order action makes a contribution to the four-point function in this limit.
Unlike that from contact diagram, the amplitude of four-point function from the exchange diagram goes like $1/{k_{12}^3}$ which is divergent in the limit of $k_{12}\rightarrow 0$. This behavior is similar to the local-form four-point function of scalar perturbation from exchange diagram.

%So unlike the case of contact diagram which doesn't diverge, exchange diagram is 3-order divergency as $k_{12}\rightarrow 0$. Its physical explanation is that for exchange diagram there is a graviton exchanged in internal line, whose spectrum is proportional to $1/{k_{12}^3}$. Similar to the ``bilateral" squeezed limit, only $c_2$ term contributes the ``diamond" squeezed limit. And the form of this limit is the square of squeezed limit of $3$-point function, see \eqref{eqn:3ql}.

\section{Summary and discussion}
\label{section:dis}

In this paper we dig out some general properties of the $n$-point function of graviton due to the parity, and explicitly compute the four-point correlation function of graviton, including both the contact and exchange diagrams, in the de Sitter approximation in the framework of GR. One can expect that the the simplicity of the results presented in this paper is lost when we go to an inflationary background.  However, our results are still the leading approximation in the slow-roll expansion. In the framework of GR, we find $\langle hhhh \rangle / \langle hh \rangle^3\sim {\cal O}(1)$.
Furthermore, the four-point function of graviton in the de Sitter background has similar behavior to the local-form four-point function of scalar perturbation in some special momentum configurations.

The three-point function of graviton has been extended to the inflationary background in \cite{Soda:2011am}, generalized G-inflation in \cite{Gao:2011vs} and Horava-Lifshitz gravity in \cite{Zhu:2013fja}. It is worthy coming back to the four-point function of graviton in more general scenario in the near future.

\vspace{5mm}
\noindent{\large \bf Acknowledgments}

We would like to thank Xian Gao, Gary Shiu and Sai Wang for useful conversations.
This work is supported by the project of Knowledge Innovation Program of Chinese Academy of Science and grants from NSFC (grant NO. 11322545 and 11335012).

%\newpage
%%%%%%%%%%%%%%%%

\appendix

\section{A brief review of tensor power spectrum and three-point correlation function of graviton in de Sitter background}
\label{app:3pt}

Following the standard technique in quantum field theory, from Eqs.~\eqref{modefunction}, \eqref{lphsk} and \eqref{lpcommu-rela}, one can easily get
\e
\langle h^{s_1}_{\mathbf{k}_1}h^{s_2}_{\mathbf{k}_2} \rangle =
 (2\pi)^3\delta^{(3)}(\mathbf{k}_1+\mathbf{k}_2)\delta_{s_1s_2}\left(\frac{H}{\Mp}\right)^2\frac{1}{2k_1^3}~.
\q
Considering the normalisation of the polarization tensor and summing over the two degrees of freedom of gravitational waves in GR, the amplitude of tensor power spectrum is given by $P_T=2(H^2/\Mp)^2/\pi^2$ which is the same as that in the literatures.

The three-point correlation function of graviton in terms of circularly polarized states in de Sitter background has been calculated in \cite{Maldacena:2002vr}. Here we adopt the linearly polarized  states and distinguish the different contributions from two interaction terms denoted by $c_1$ and $c_2$ in the third order action \eqref{3rd-action}. Using the the first order in-in formalism, we obtain
 \e
\langle h^{s_1}_{\mathbf{k}_1}h^{s_2}_{\mathbf{k}_2}h^{s_3}_{\mathbf{k}_3}\rangle = (2\pi)^3 \delta^3 (\sum_{i=1}^3\mathbf{k}_i) \left(\frac{H}{\Mp}\right)^4\frac{\mathcal{A}^{s_1 s_2 s_3}(\mathbf{k}_1,\mathbf{k}_2,\mathbf{k}_3)}{(k_1k_2k_3)^3}~,
\q
where
\m
\mathcal{A}^{s_1 s_2 s_3}(\mathbf{k}_1,\mathbf{k}_2,\mathbf{k}_3) &=&\frac{1}{16} \left(-\mathcal{K} + \frac{1}{\mathcal{K}} \sum_{i<j} k_i k_j+\frac{k_1k_2k_3}{\mathcal{K}^2}\right) \nn \\
 &\times& \left[(c_1 k_3^k e^1_{ki}e^3_{ij}e^2_{jl}k_3^l-\frac{c_2}{2}e^1_{ij}e^3_{ij}k_3^k e^2_{kl}k_3^l) + \text{5 perms.}\right]~,
\n
and $\mathcal{K}=k_1+k_2+k_3$.
The energy-momentum conservation implies that all the momenta $\mathbf{k}_i$ in the three-point function must lie in the same plane and then only the amplitudes of three-point function with even number of $\times$ polarized states do not equal zero. Taking the permutations into account, there are only two independent non-zero components
\m
\mathcal{A}^{\times\times+}(\mathbf{k}_1,\mathbf{k}_2,\mathbf{k}_3)&=&\frac{(4c_1k_3^2+ c_2 (k_1^2+k_2^2-k_3^2))}{16\sqrt{2}k_1 k_2 k_3^2\cal{K}}
\(-k_1 k_2 k_3 -{\cal{K}} \sum_{i<j}k_i k_j +{\cal K }^3 \) \nn\\
&&\times \(8 k_1 k_2 k_3 -4{\cal{K}} \sum_{i<j}k_i k_j +{\cal K }^3 \), \\
\mathcal{A}^{+++}(\mathbf{k}_1,\mathbf{k}_2,\mathbf{k}_3)&=& \frac{1}{64\sqrt{2}(k_1 k_2 k_3)^2\mathcal{K}}
\(8 k_1 k_2 k_3 - 4 \mathcal{K}\sum_{i<j}k_i k_j +\mathcal{K}^3\) \nn\\
&&\times\(k_1 k_2 k_3+\mathcal{K}\sum_{i<j}k_i k_j-\mathcal{K}^3\) \nn \\
&&\times \left[8c_2 (\sum_{i<j}k_i k_j)^2-8(2c_1-c_2)\mathcal{K}k_1 k_2 k_3 \right. \nn\\
&& \left. +4(2c_1-3c_2)\mathcal{K}^2\sum_{i<j}k_i k_j+(3c_2-2c_1)\mathcal{K}^4\right].
\n
In the squeezed limit, i.e. $k_1=k_2=k$ and $k_3\rightarrow0$, we have
\e\label{eqn:3ql}
\mathcal{A}^{s_1 s_2 s_3}= \frac{3c_2 k^3}{16\sqrt{2}}(s_1+s_2)(1+s_3),
\q
which says that only the interaction term denoted by $c_2$ makes contributions to the three-point function in the squeezed limit. Note that $s=+1,\ -1$ for $+,\ \times$ polarized states respectively. From the above formula, there are only two non-zero components, i.e.
\e
\mathcal{A}^{+++}=-\mathcal{A}^{\times\times+}= \frac{3c_2 k^3}{4\sqrt{2}},
\q
in the squeezed limit $(k_3\rightarrow 0)$. It is similar to the local-form three-point function of scalar perturbation during inflation.

\section{Polarization tensors}
\label{app:poltensor}

\begin{figure*}[htbp]
\centering
\begin{center}
\includegraphics[scale=0.8]{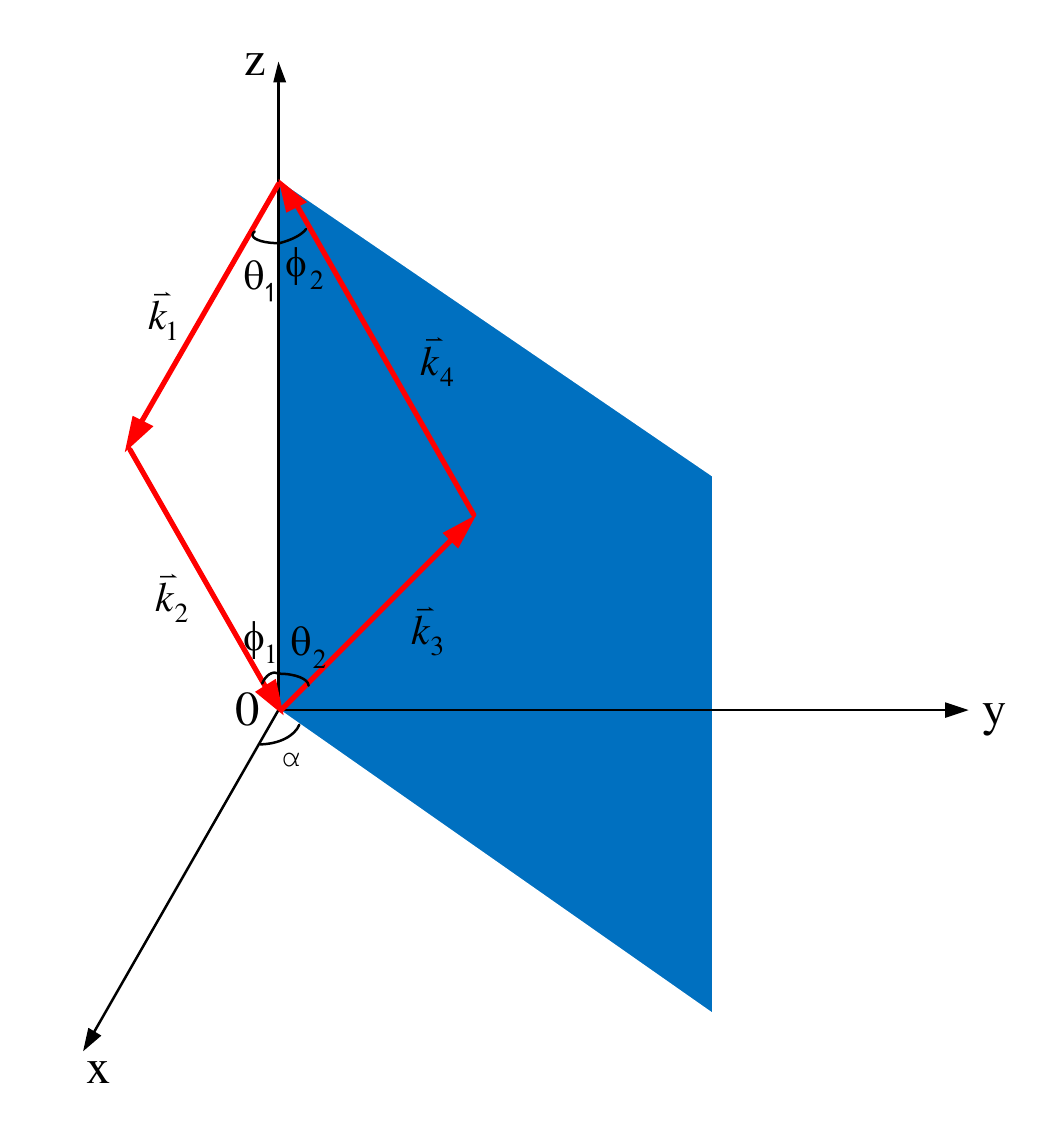}
\end{center}
\caption{Momentum configuration. }
\label{figpolarization}
\end{figure*}

In this section, we will fix the representation of polarization tensors. Due to the momentum conservation
 $\mathbf{k}_1+\mathbf{k}_2+\mathbf{k}_3+\mathbf{k}_4=0$ for the four-point function, the momentum configuration can be described by six parameters, i.e. $k_{12}$, $\theta_1$, $\theta_2$, $\phi_1$, $\phi_2$ and $\alpha$, which are illustrated in Fig.~\ref{figpolarization}.
Here, without loss of generality, we assume that $\mathbf{k}_1$ and $\mathbf{k}_2$ lie in the $x$-$z$ plane, and $-\mathbf{k}_{12}$ stays along $\hat{\mathbf{z}}=(0,0,1)^{\mathrm{T}}$ direction. And then $\mathbf{k}_i$ can be expressed by
\m
 \mathbf{k}_1&=&k_1R_2(\pi-\theta_1)\hat{\mathbf{z}},\\
 \mathbf{k}_2&=&k_2R_2(\pi+\phi_1)\hat{\mathbf{z}},\\
 \mathbf{k}_3&=&k_3R_3(\alpha)R_2(\theta_2)\hat{\mathbf{z}},\\
 \mathbf{k}_4&=&k_4R_3(\pi+\alpha)R_2(\phi_2)\hat{\mathbf{z}},
\n
where $R_i(\theta_i)$ denotes rotating $\theta_i$ about $x_i$ axis:
\m  \label{transmatrix}
R_1(\theta_1)&=&
\left(
  \begin{array}{ccc}
    1 & 0 & 0\\
    0 & \cos\theta_1 & -\sin\theta_1\\
    0 & \sin\theta_1 & \cos\theta_1\\
  \end{array}
  \right), \\
  R_2(\theta_2)&=&
\left(
  \begin{array}{ccc}
    \cos\theta_2 &0 & \sin\theta_2\\
    0          & 1 & 0\\
    -\sin\theta_2 & 0 & \cos\theta_2\\
  \end{array}
  \right), \\
  R_3(\theta_3)&=&
\left(
  \begin{array}{ccc}
    \cos\theta_3 & -\sin\theta_3 & 0\\
    \sin\theta_3 & \cos\theta_3 & 0\\
    0 & 0 & 1\\
  \end{array}
  \right).
 \n
Now let's start with the linear polarization tensor along $\hat{\mathbf{z}}$
\e
e^{s}_{ij}(\hat{\mathbf{z}})=\sqrt{2}
 \left(
  \begin{array}{ccc}
    \frac{1+s}{2} & \frac{1-s}{2} & 0\\
    \frac{1-s}{2} & -\frac{1+s}{2} & 0\\
     0 & 0 & 0\\
  \end{array}
  \right),
\q
where $s=1\ (-1)$ for $+\ (\times)$ polarization state respectively.
And then the polarization tensors along $\mathbf{k}_i$ directions are given by
\m
e^{s}(\mathbf{k}_1)&=& R_2(\pi-\theta_1)e^{s}(\hat{\mathbf{z}}){R_2(\pi-\theta_1)}^{\mathrm{T}}, \\
e^{s}(\mathbf{k}_2)&=& R_2(\pi+\phi_1)e^{s}(\hat{\mathbf{z}}){R_2(\pi+\phi_1)}^{\mathrm{T}}, \\
e^{s}(\mathbf{k}_3)&=& R_3(\alpha)R_2(\theta_2)e^{s}(\hat{\mathbf{z}}){R_2(\theta_2)}^{\mathrm{T}}{R_3(\alpha)}^{\mathrm{T}}, \\
e^{s}(\mathbf{k}_4)&=& R_3(\pi+\alpha)R_2(\phi_2)e^{s}(\hat{\mathbf{z}}){R_2(\phi_2)}^{\mathrm{T}}{R_3(\pi+\alpha)}^{\mathrm{T}}.
\n

\section{The constraints on the correlation function of graviton from the de Sitter isometries}
\label{app:isometries}

There are $10$ isometries in the four-dimensional de Sitter spacetime, including 3 spatial rotations, 3 spatial translations, 1 dilatation and 3 special conformal transformations (SCT). For de sitter space, the spatial rotations and translations are trivial symmetries which just indicates isotropy and homogeneity.

In the conformal frame, $ds^2=(-d\eta^2+dx^2)/\eta^2$ and there is a scaling symmetry under the transformation $\eta\rightarrow \lambda\eta$, $\mathbf{x}\rightarrow\lambda \mathbf{x}$. Therefore we have
\m
\langle h_{i_1 j_1}^{s_1}(\mathbf{x}_1,\eta)...h_{i_n j_n}^{s_n}(\mathbf{x}_1,\eta)\rangle=\langle h_{i_1 j_1}^{s_1}(\lambda\mathbf{x}_1,\lambda\eta)...h_{i_n j_n}^{s_n}(\lambda\mathbf{x}_2,\lambda\eta)\rangle~,
\n
which reads
\m
\langle h^{s_1}_{\mathbf{k}_1}(\eta)... h^{s_n}_{\mathbf{k}_n}(\eta)\rangle=\frac{1}{\lambda^{3n}}\langle  h^{s_1}_{\mathbf{k}_1 /\lambda}(\lambda\eta)...h^{s_n}_{\mathbf{k}_n /\lambda }(\lambda\eta)\rangle~
\n
for the Fourier modes.
From Eq.~(\ref{Tnp}), taking the late time limit (i.e. $\eta\rightarrow0$), we find
\m
\mathcal{T}^{s_1  \cdots s_n}(\mathbf{k}_1,\cdots,\mathbf{k}_n)=\frac{1}{\lambda^{3(n-1)}}\mathcal{T}^{s_1  \cdots s_n}(\mathbf{k}_1/\lambda,\cdots,\mathbf{k}_n/\lambda),
\n
or equivalently
\m
\bigg[3(n-1)+\sum_{a=1}^{n}\mathbf{k}_a\cdot\partial_{\mathbf{k}_a}\bigg]T(\mathbf{k}_1,...,\mathbf{k}_n)=0.
\n
The dilatation invariance implies that the spectrum of gravity fluctuations is scale-invariant.

However, the SCT, namely $\eta\rightarrow \eta-2\eta(\mathbf{b}\cdot\mathbf{x})$ and $\mathbf{x}\rightarrow \mathbf{x}+\mathbf{b}(-\eta^2+\mathbf{x}^2)-2\mathbf{x}(\mathbf{b}\cdot\mathbf{x})$, where $\mathbf{b}$ is infinitesimal, changes the spatial slices and then violates the ADM decomposition. Even though in the late time limit SCT becomes
 $\eta\rightarrow \eta$ and $\mathbf{x}\rightarrow \mathbf{x}+\mathbf{b}(\mathbf{x}^2)-2\mathbf{x}(\mathbf{b}\cdot\mathbf{x})$, which doesn't change the spatial slices, the gravity fluctuation $h_{ij}$ breaks the traceless and transverse conditions under this transformation \cite{Hinterbichler:2012}. In this sense, SCT is not a good symmetry for gravitational waves in our gauge-fixing 
 system. Even though SCT may provide a constraint on the structure of the $n$-point function of graviton in the de Sitter background, how to work it out is still an open question. 
 
These four isometries are absent in the realistic inflationary background and the structures of the $n$-point function in the de Sitter background should be corrected up to the slow-roll parameters.


\begin{thebibliography}{99}


\bibitem{Riess:1998cb}
  A.~G.~Riess {\it et al.}  [Supernova Search Team Collaboration],
  %``Observational evidence from supernovae for an accelerating universe and a cosmological constant,''
  Astron.\ J.\  {\bf 116}, 1009 (1998)
  [astro-ph/9805201].

\bibitem{Perlmutter:1998np}
  S.~Perlmutter {\it et al.}  [Supernova Cosmology Project Collaboration],
  %``Measurements of Omega and Lambda from 42 high redshift supernovae,''
  Astrophys.\ J.\  {\bf 517}, 565 (1999)
  [astro-ph/9812133].


\bibitem{Starobinsky:1980te}
  A.~A.~Starobinsky,
  %``A New Type of Isotropic Cosmological Models Without Singularity,''
  Phys.\ Lett.\ B {\bf 91}, 99 (1980).

\bibitem{Guth:1980zm}
  A.~H.~Guth,
  %``The Inflationary Universe: A Possible Solution to the Horizon and Flatness Problems,''
  Phys.\ Rev.\ D {\bf 23}, 347 (1981).

\bibitem{Linde:1981mu}
  A.~D.~Linde,
  %``A New Inflationary Universe Scenario: A Possible Solution of the Horizon, Flatness, Homogeneity, Isotropy and Primordial Monopole Problems,''
  Phys.\ Lett.\ B {\bf 108}, 389 (1982).

\bibitem{Albrecht:1982wi}
  A.~Albrecht and P.~J.~Steinhardt,
  %``Cosmology for Grand Unified Theories with Radiatively Induced Symmetry Breaking,''
  Phys.\ Rev.\ Lett.\  {\bf 48}, 1220 (1982).


\bibitem{Maldacena:1997re}
  J.~M.~Maldacena,
  %``The Large N limit of superconformal field theories and supergravity,''
  Int.\ J.\ Theor.\ Phys.\  {\bf 38}, 1113 (1999)
  [Adv.\ Theor.\ Math.\ Phys.\  {\bf 2}, 231 (1998)]
  [hep-th/9711200].



\bibitem{Starobinsky:1979}
A.~A.~Starobinsky,
  %``Spectrum of relict gravitational radiation and the early state of the
  %universe,''
  JETP Lett.\  {\bf 30}, 682 (1979)
  [Pisma Zh.\ Eksp.\ Teor.\ Fiz.\  {\bf 30}, 719 (1979)].


\bibitem{Maldacena:2002vr}
  J.~Maldacena,
  ``Non-Gaussian features of primordial fluctuations in single field
  inflationary models,''
  JHEP {\bf 0305}, 013 (2003)
  [arXiv:astro-ph/0210603].


\bibitem{Maldacena:2011nz}
  J.~M.~Maldacena and G.~L.~Pimentel,
  %``On graviton non-Gaussianities during inflation,''
  JHEP {\bf 1109}, 045 (2011)
  [arXiv:1104.2846 [hep-th]].



\bibitem{Arnowitt:1962hi}
  R.~L.~Arnowitt, S.~Deser and C.~W.~Misner,
  %``The Dynamics of general relativity,''
  Gen.\ Rel.\ Grav.\  {\bf 40}, 1997 (2008)
  [gr-qc/0405109].


\bibitem{Bunch:1978yq}
  T.~S.~Bunch and P.~C.~W.~Davies,
  ``Quantum Field Theory in de Sitter Space: Renormalization by Point Splitting,''
  Proc.\ Roy.\ Soc.\ Lond.\ A {\bf 360}, 117 (1978).



\bibitem{Weinberg:2005vy}
  S.~Weinberg,
  ``Quantum contributions to cosmological correlations,''
  Phys.\ Rev.\ D {\bf 72}, 043514 (2005)
  [hep-th/0506236].



\bibitem{Soda:2011am}
  J.~Soda, H.~Kodama and M.~Nozawa,
  %``Parity Violation in Graviton Non-gaussianity,''
  JHEP {\bf 1108}, 067 (2011)
  [arXiv:1106.3228 [hep-th]].

\bibitem{Gao:2011vs}
  X.~Gao, T.~Kobayashi, M.~Yamaguchi and J.~Yokoyama,
  %``Primordial non-Gaussianities of gravitational waves in the most general single-field inflation model,''
  Phys.\ Rev.\ Lett.\  {\bf 107}, 211301 (2011)
  [arXiv:1108.3513 [astro-ph.CO]].

\bibitem{Zhu:2013fja}
  T.~Zhu, W.~Zhao, Y.~Huang, A.~Wang and Q.~Wu,
  %``Effects of parity violation on non-gaussianity of primordial gravitational waves in Ho\v{r}ava-Lifshitz gravity,''
  Phys.\ Rev.\ D {\bf 88}, 063508 (2013)
  [arXiv:1305.0600 [hep-th]].

\bibitem{Hinterbichler:2012}
 K.~Hinterbichler,L.~Hui,J.~Khoury,
 JCAP, 2012(08), 017.

\end{thebibliography}
\end{document}